\documentclass[lettersize,journal]{IEEEtran}

\usepackage{algorithmic}
\usepackage{algorithm}
\usepackage{amsmath,amsfonts}
\usepackage{algorithmic}
\usepackage{algorithm}
\usepackage{array}
\usepackage[caption=false,font=normalsize,labelfont=rm,textfont=rm]{subfig}
\usepackage{textcomp}
\usepackage{stfloats}
\usepackage{url}
\usepackage{verbatim}
\usepackage{graphicx}
\usepackage{cite}
\usepackage{amsthm}
\usepackage{acronym}
\usepackage{multirow}
\usepackage{makecell}
\usepackage[hidelinks]{hyperref}
\usepackage{eso-pic}

\DeclareMathAlphabet\mathbfcal{OMS}{cmsy}{b}{n}

\acrodef{AI}[AI]{Artificial Intelligence}
\acrodef{IoT}[IoT]{Internet of Things}
\acrodef{DNN}[DNN]{Deep Neural Network}
\acrodef{M2M}[M2M]{machine-to-machine}
\acrodef{MI}[MI]{mutual information}
\acrodef{DPCM}[DPCM]{Differential Pulse Width Modulation}
\acrodef{DCT}[DCT]{Discrete Cosine Transform}
\acrodef{DP}[DP]{Differential Privacy}
\acrodef{QP}[QP]{Quantization Parameter}
\acrodef{MIA}[MIA]{Model Inversion Attack}
\acrodef{VVC}[VVC]{Versatile Video Coding}
\acrodef{GAN}[GAN]{Generative Adversarial Network}

\newtheorem{lemma}{Lemma}

\begin{document}

\AddToShipoutPicture*{\small \sffamily\raisebox{27.0cm}{\hspace{1.6cm}
 To be Published in IEEE Transactions on Image Processing}}

\title{Privacy-Preserving Autoencoder for Collaborative Object Detection}

\author{Bardia Azizian and Ivan V. Baji\'{c}

\thanks{This work was supported in part by a gift from Intel Labs, NSERC grants RGPIN-2021-02485 and RGPAS-2021-00038, and Digital Research Alliance of Canada.}
\thanks{The authors are with the School of Engineering Science, Simon Fraser University, Burnaby, BC V5A 1S6 Canada (e-mail: bardia\_azizian@sfu.ca; ibajic@ensc.sfu.ca)}
}

\maketitle

\begin{abstract}
Privacy is a crucial concern in collaborative machine vision where a part of a \ac{DNN} model runs on the edge, and the rest is executed on the cloud. In such applications, the machine vision model does not need the exact visual content to perform its task. Taking advantage of this potential, private information could be removed from the data insofar as it does not significantly impair the accuracy of the machine vision system. In this paper, we present an autoencoder-style network integrated within an object detection
pipeline, which generates a latent representation of the input image that preserves task-relevant information while removing private information. Our approach employs an adversarial training strategy that not only removes private information from the bottleneck of the autoencoder but also promotes improved compression efficiency for feature channels coded by conventional codecs like VVC-Intra.
We assess the proposed system using a realistic evaluation framework for privacy, directly measuring face and license plate recognition accuracy. Experimental results show that our proposed method is able to reduce the bitrate significantly at the same object detection accuracy compared to coding the input images directly, while keeping the face and license plate recognition accuracy on the images recovered from the bottleneck features low, implying strong privacy protection. Our code is available at \href{https://github.com/bardia-az/ppa-code}{https://github.com/bardia-az/ppa-code}.
\end{abstract}

\begin{IEEEkeywords}
Deep neural network, coding for machines, privacy, model inversion attack, collaborative intelligence, adversarial training, feature compression
\end{IEEEkeywords}

\IEEEpeerreviewmaketitle

\section{Introduction}
\IEEEPARstart{W}{ith} recent advancements in \ac{AI} applications such as autonomous driving, visual surveillance, and \ac{IoT}, the volume of data being transmitted between ``intelligent'' edge devices (e.g., cameras) and the cloud-based services is rapidly increasing. In these applications, the inference is performed collaboratively between an edge device, which often has limited computational capabilities, and a more powerful computing entity commonly referred to as the cloud.
Given the rapidly growing number of such \ac{M2M} connections~\cite{cisco_report}, there is an urgent need for an efficient data compression methodology tailored to automated machine-based analysis.
JPEG AI~\cite{JPEG-AI, JPEG_AI_paper} and MPEG-VCM~\cite{MPEG-VCM} are two prominent standardization groups actively working in this area to address this pressing need for data compression for machines~\cite{VCM_paper}. As part of our research objectives in this work, we also aim to improve compression efficiency for such collaborative \ac{M2M} analysis systems.

At the same time, privacy has always been a major concern in data communications. This concern becomes increasingly critical as the volume of data collected and passed around among edge devices and the cloud continues to grow. Consider a scenario where an adversary gains unauthorized access to data collected by traffic cameras and other sensors used in traffic monitoring. If they are able to identify individuals from such data, they could track them, determine their lifestyle patterns, and then use such knowledge for a variety of illegal and harmful purposes. Similar concerns exist in smart home applications, independent living, etc.
Cryptography schemes such as data encryption~\cite{cryptography_cst2016} or secret sharing~\cite{secret_share, SIFT_outsource} offer one possible solution for protecting data, although they come with their own risks and challenges, like the need for a trusted party.

Collaborative intelligence~\cite{kang2017neurosurgeon} (also referred to as collaborative inference) has the potential to mitigate the risk of exposing private information, as data passed between the edge device and the cloud are not the original input data, but some intermediate features of a \ac{DNN} model performing the intended analysis. In this framework, the first few layers of a \ac{DNN} model are executed on the edge, and the resulting features are coded and transmitted to the cloud, where the remaining layers are processed. This setup resembles federated learning~\cite{federated_learning} and split learning~\cite{SplitFed}, where the training process is distributed across multiple decentralized entities, usually to protect the privacy of training data. In collaborative intelligence, however, the inference is carried out in a decentralized manner. Hence, protecting data privacy during inference is of particular interest in such a scenario.

\begin{figure}[!t]
    \centering
    \includegraphics[width=\columnwidth]{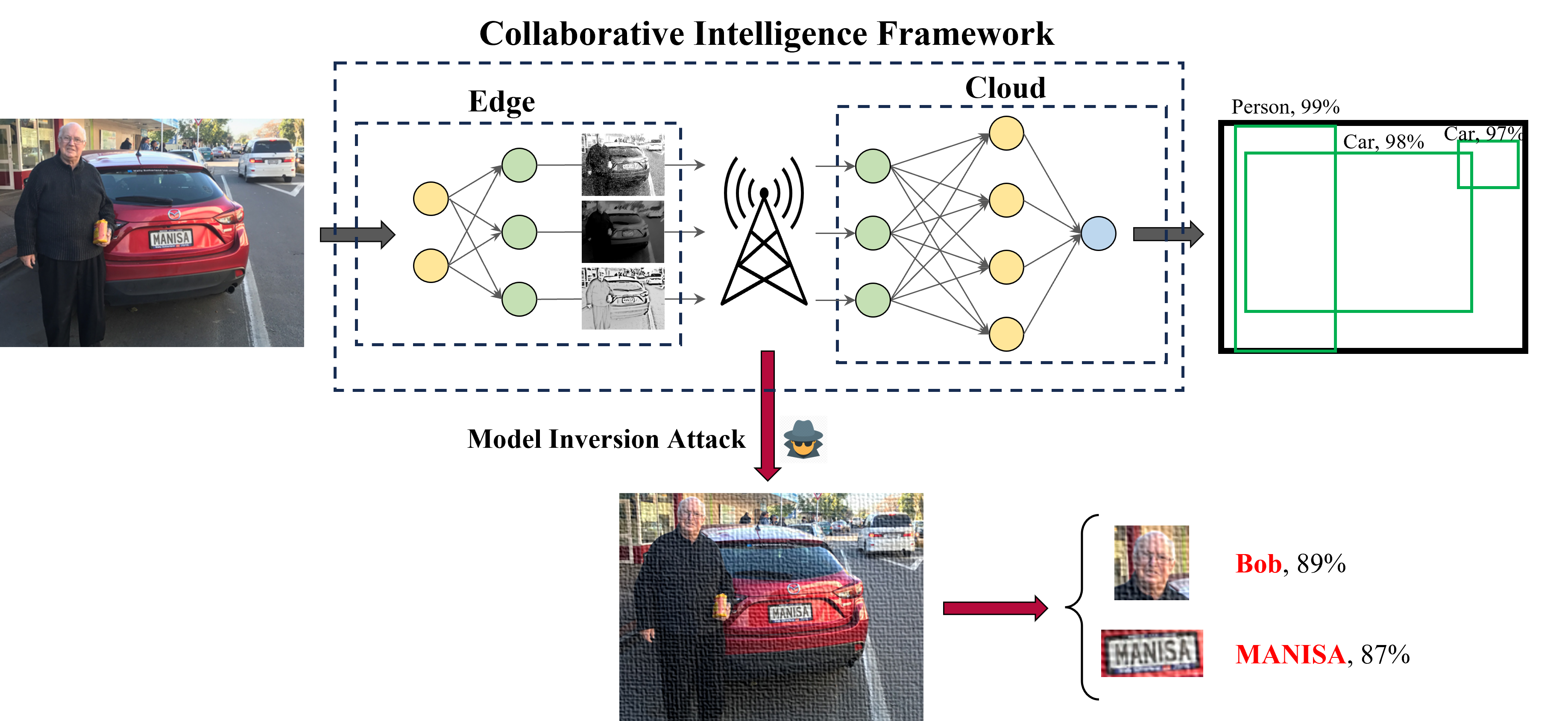}
    \caption{An object detection model deployed in a collaborative intelligence framework with an adversary performing a model inversion attack.}
    \label{fig:collab_intl}
    \vspace{-0.4cm}
\end{figure}

Distributing the computations of \acp{DNN} among edge and cloud in collaborative intelligence can bring latency and power efficiency \cite{Collab_eshratifar, Collab_eshratifar_2, Collab_Auto-Split}. 
Although this approach reveals only a latent representation of the input data, it remains vulnerable to the \ac{MIA}~\cite{mia_2019} through which the input data can be recovered to some extent, as illustrated in Fig.~\ref{fig:collab_intl}. MIA is a very general privacy attack and is not related to any particular notion of privacy. Instead, its goal is to recover the input signal which, if successful, could lead to inferring any private information contained in the input, even the kind that is yet to be recognized as privacy-sensitive, for example, the likelihood of dementia. Hence, resistance to MIA provides fairly general privacy protection.
To counter this attack, one strategy would be to remove task-irrelevant information -- and, in doing so, private information as well -- while retaining only task-relevant information in the communicated data. However, distinguishing between task-relevant, task-irrelevant, and private information remains a challenge, especially in real-world visual data.

This paper is an extension of our previous study~\cite{bardia_pcs2022} and presents the following novel contributions:
\begin{enumerate}
    \item Addressing both compression efficiency and privacy simultaneously: Data compression and privacy have traditionally been considered two distinct subjects in the literature. In this paper, however, we propose a practical framework with improved privacy and compression efficiency, which could be deployed in a collaborative intelligence setting.
    \item Deeper theoretical analysis of the proposed approach: We analyze \ac{MIA} from an information-theoretic perspective, highlighting some of the challenges and how our approach tackles them. 
    \item A novel adversarial training technique: The proposed training procedure and the loss functions are designed in a way to encourage higher compression efficiency and machine vision accuracy while simultaneously discouraging precise input reconstruction from transmitted data.
    \item A new evaluation framework for privacy assessment: Most \ac{MIA}-related research, including our earlier work~\cite{bardia_pcs2022}, uses perceptual quality metrics as proxy criteria for privacy. In contrast, our proposed evaluation framework takes personal identity as private information and examines how well an adversary can infer personal identity from images containing faces or license plates.\footnote{License plates can be associated with personal identity through vehicle license records.}
\end{enumerate}

The remainder of the paper is organized as follows. In Section~\ref{sec:rel_work}, a review of the related work is provided. Section~\ref{sec:prelim} presents the preliminaries needed for subsequent material, as well as information-theoretic analysis and the resulting motivation for the proposed approach. Sections~\ref{sec:method} and~\ref{sec:eval_frame} present our proposed methods and evaluation framework. The experimental results are given in Section~\ref{sec:exp_res}, followed by the conclusions in Section~\ref{sec:conclusion}.

\section{Related Work}
\label{sec:rel_work}
For decades, image and video codecs~\cite{rabbani_overview_2002, HEVC_overview, VVC_overview} have evolved with a focus on reducing the number of bits representing input data while maintaining fidelity between the original and reconstructed input. The latter has been pursued by improving perceptual quality metrics~\cite{Lin_Kuo_JVCIR_2011}, as the end users have traditionally been humans. However, in emerging applications such as traffic monitoring, smart homes, etc., visual content is almost exclusively ``looked at'' by machines, with only occasional, if any, human viewing.
In such cases, coding only the task-relevant features derived from the input is more advantageous than (auto-)encoding the input itself, in terms of both compression efficiency and privacy. Even prior to the current wave of interest in \acp{DNN}, the significance of feature coding had been recognized through the development of MPEG standards such as Compact Descriptors for Visual Search (CDVS)~\cite{cdvs_std} and Compact Descriptors for Visual Analysis (CDVA)~\cite{cdva_std}. More recently, feature coding has been studied in the context of collaborative intelligence~\cite{CI_ICASSP_2021, collab_int_new}, as well as through standardization efforts such as MPEG-VCM~\cite{MPEG-VCM}, where a major goal is to develop efficient compression methods for feature tensors of intermediate \ac{DNN} layers.

Another related line of research has been the use of \acp{DNN} for image coding, with early examples found in the work of Ball\'{e} \textit{et al.}~\cite{balle_end_to_end} and subsequent work~\cite{balle_hyperprior, balle_autoregressive, Cheng_2020_CVPR, ANFIC, video_survey1}. Latest-generation \ac{DNN}-based image codecs~\cite{Elic, MLIC, MLIC++, LIC_att, LIC_GLLMM, Chen_LIC} have surpassed state-of-the-art hand-crafted codecs such as \ac{VVC}-Intra. Although \ac{DNN}-based image codecs are typically trained using quality metrics such as Peak Signal-to-Noise Ratio (PSNR) and Structural Similarity Index Metric (SSIM)~\cite{SSIM_TIP2004}, which are targeted at human viewing, they can also be trained using machine task-relevant loss functions~\cite{End_learned_machines, End_learned_machines_2}. Nonetheless, so long as the main goal is to perform a machine vision task, encoding and decoding the entire input image seems unnecessary.
Instead, it is preferable to run a part of the machine vision \ac{DNN} model at the edge and transmit the resulting machine-targeted features to the cloud. These features can sometimes be used for multiple tasks on the cloud~\cite{Saeed_pareto}.

Like natural images, \ac{DNN} features can be coded using either \ac{DNN}-based~\cite{YOLOv5_based_coding} or hand-crafted methods~\cite{Hyomin_deep_feature_compression, Collab_near_lossless, bottlenet, Bob_Collab_Clip}. The main advantage of \ac{DNN}-based codecs is their flexibility, which enables them to be trained end-to-end with a loss function that combines various objectives. Leveraging this flexibility can further lead to scalable coding for multiple downstream tasks~\cite{scalable_coding_Hyomin, SSSIC}.
On the other hand, traditional codecs may still be preferable in some cases for reasons such as widespread availability in hardware and software, as well as user trust developed and time-tested over the years.
However, their main drawback is that they have been optimized to encode natural images, which could lead to suboptimal performance when encoding \ac{DNN} features. A solution to alleviate this problem is to tailor \ac{DNN} features to align with the utilized codec through fine-tuning the \ac{DNN} model. This can be achieved by introducing appropriate differentiable proxies for the traditional codecs~\cite{sandwich2021, diff_bitrate, sandwich2022}, thereby optimizing the entire system in an end-to-end manner. Similarly,~\cite{compressible_loss} has proposed a simpler differentiable proxy for the bitrate of a traditional codec that takes into account its basic processing steps.
We adopted this approach in our proposed method to enhance the compressibility of features by the VVC codec.

As mentioned above, the field of image coding has not yet embraced privacy considerations in a substantial way. However, researchers in the field of machine/deep learning have been addressing privacy for some time~\cite{privacy_survey_2, privacy_survey_1, privacy_survey_3}. Some focus on protecting the data used during the training phase~\cite{learn_pp_student}, while others seek data privacy during inference, which is also the objective of this article. Various privacy and security attacks exist, along with corresponding defense mechanisms, for deep learning models, based on the type of privacy concern. In this paper, we focus on the defense against \ac{MIA}. In the \ac{MIA}, the adversary aims to recover the input image to the \ac{DNN} model either during the training or inference phase, by observing only the outputs or some intermediate features of the model. For example,~\cite{mia_face_recovery} introduced an attack in which the image of a person used in training could be recovered through access to the facial recognition \ac{DNN} model, its confidence score output, and knowledge of the person's name. The authors of~\cite{mia_2019} also attempt to recover the input image to a classification \ac{DNN} model by observing the intermediate features during inference under different conditions and the adversary's level of access to the model.

While the data privacy of the \ac{DNN} models during the inference phase is less explored, the privacy of the models and their training data has been on the radar of researchers for many years. The concept of \ac{DP}~\cite{diff_privacy} was introduced to protect the private information of individuals contributing to a dataset utilized by a machine learning or analysis model. The main idea in \ac{DP} is adding noise to the output of the deterministic analysis model to increase uncertainty about the dataset, such that distinguishing between two datasets differing in only one entity becomes nearly impossible. Some research works have employed \ac{DP} to protect neural networks against input reconstruction performed by \ac{MIA}~\cite{diff_privacy_iot, not_just_privacy, pricure, mia_2021}. For instance, in~\cite{not_just_privacy},  Laplacian noise is added to the intermediate features of the \ac{DNN} model, which are going to be transmitted to the cloud.
This would randomize the deterministic operations within the \ac{DNN} model and increase the uncertainty in the input given the intercepted features. Additionally, a nullification operation is also conducted to enhance privacy. To increase the robustness of the cloud-side back-end to such feature perturbations, it is retrained on the manipulated features. However, the utility of the model is severely affected by increasing the intensity of the noise and nullification process. The same methodology is adopted in~\cite{diff_privacy_iot}, where the Privacy-Utility trade-off is studied from the \ac{DP} perspective. This trade-off can be controlled by the privacy budget which has an inverse relationship with the Laplacian noise distribution's scale parameter.

The authors of \cite{mia_2019} have extended their research in a recent paper~\cite{mia_2021}, by offering defense strategies against their proposed types of \acp{MIA} introduced in \cite{mia_2019}. They investigated noise obfuscation using Laplacian and Gaussian noise, as well as the dropout technique. Among these, the dropout method, which involves the deactivation of random neurons within a \ac{DNN} layer, demonstrated substantially better performance in terms of the Privacy-Utility trade-off.

In the existing literature, there is related work whose goal is to find a transformation of the input images to eliminate sensitive private attributes while preserving other information for potential analysis. Various methods, including anonymization and de-identification techniques, have been explored. Early approaches primarily relied on traditional schemes such as filtering, pixelation, blurring, or superpixel clustering~\cite{trad_anonym}. More recent methods leverage \acp{GAN}~\cite{GAN} to create realistic and visually pleasant substitutes for removed sensitive regions~\cite{Maximov_2020_CVPR, Auto-Driving_GAN, sec_id_anonym, wu_privacy-protective-gan_2019}. However, these techniques require prior knowledge of the location of sensitive regions, posing a challenge for edge devices with constrained computational resources. In such methods, the discriminator's job is to detect whether its input image is produced by the generator or is real, similar to typical \ac{GAN} frameworks. In contrast, in studies like~\cite{adv_privacy_1, Wu_2018_ECCV_privacy, adv_privacy_2}, the discriminator assumes the role of an adversary seeking to infer private information from the generator's output representation. Here, the generator competes with a privacy-attacking discriminator throughout the training and learns to remove sensitive information that could enable inference of private attributes. This resembles our proposed adversarial training, but with two important differences: (1) instead of a generator, we use a much simpler autoencoder, which is more suitable for edge device deployment, and (2) we focus on compression efficiency, which is essential for edge-cloud communication.

The main vision tasks employed in~\cite{adv_privacy_1} and~\cite{Wu_2018_ECCV_privacy} are classification and action recognition, respectively. In both cases, the discriminator is a classification model, aiming to predict specific classes or private attributes. 
The authors of~\cite{adv_privacy_1} mainly focus on improving the stability of the adversarial training process by imposing constraints like per-location normalization after each layer of the generator to avoid collapsing to a degenerate solution with constant outputs. On the other hand, the main goal in~\cite{Wu_2018_ECCV_privacy}, is to enhance the generator's generalization to unseen attacker models by frequently resetting the discriminator's weights to random values throughout the training, thereby preventing the generator from trivially overfitting to a specific discriminator. The application of~\cite{adv_privacy_2} is in medical images, where a Siamese discriminator is utilized to determine if its input pair of images belongs to the same person. This way, the generator learns to obfuscate the identity-related information while preserving task-specific details. However, our proposed method adopts a more robust discriminator capable of conducting~\acp{MIA} to recover the whole image. Consequently, our autoencoder (which plays a similar role as generators in the above studies) learns to remove more task-irrelevant information and improve compression efficiency.

Unlike the schemes mentioned above, \cite{Privacy_Saeed} incorporated privacy preservation into the feature compression process. Features with less private and more non-private information are lightly compressed, whereas those carrying more sensitive data undergo a more intensive compression to ensure privacy. Distinguishing between private and non-private features is done through an information-theoretic privacy approach called \textit{privacy fan}. In this approach, features of a \ac{DNN} are ranked according to their \ac{MI}~\cite{Cover_Thomas_2006} relative to private and non-private tasks. At its core, the privacy fan is a \ac{MI}-based feature selection mechanism. However, estimating \ac{MI} in a high-dimensional feature space is extremely challenging. Our proposed approach avoids this challenge, as it concentrates on feature modification rather than feature selection.

\section{Preliminaries, Analysis, and Motivation}
\label{sec:prelim}

\subsection{Collaborative Intelligence Formulation}
\label{subsec:ci_form}
A \ac{DNN} is fundamentally a parametric function with the set of parameters $\theta$, which transforms an input vector $x \in \mathcal{X}$ to an output vector $\bar{v} \in \mathcal{V}$, while $v \in \mathcal{V}$ represents the ground-truth output. 
\begin{equation}
    f(\cdot \, ; \theta): \mathcal{X} \to \mathcal{V} \\
\end{equation}
As mentioned earlier, collaborative intelligence is a framework in which the \ac{DNN} model is divided into a front-end $f_1(\cdot \, ; \theta_1)$ and a back-end $f_2(\cdot \, ; \theta_2)$, each with its set of parameters $\theta_1$ and $\theta_2$, respectively, such that:
\begin{equation}
    y = f_1(x; \theta_1); \quad \bar{v} = f_2(y; \theta_2),
\end{equation}
where $y$ represents the intermediate feature tensor. Throughout this paper, lowercase letters $x$, $v$, $\bar{v}$ and $y$ denote specific outcomes of random vectors (tensors) $\mathbf{X}$, $\mathbf{V}$, $\overline{\mathbf{V}}$, and $\mathbfcal{Y}$, respectively.

As discussed in \cite{kang2017neurosurgeon}, distributing the \ac{DNN} model between the edge and the cloud can improve latency and energy efficiency. However, choosing a suitable split point in the network is a design challenge~\cite{kang2017neurosurgeon,Collab_eshratifar_2}, which depends on many factors such as energy considerations, computational resources of the edge and cloud, and the communication channel. Here, our selection of the split point is based on considerations of privacy and compression efficiency.

Consider a trained feedforward \ac{DNN} with fixed parameters (weights) $\theta$. Let $\mathbfcal{Y}_i$ be the feature tensor at the $i$-th layer.
Then $\mathbf{X} \to \mathbfcal{Y}_i \to \mathbfcal{Y}_{i+1}$ forms a Markov chain and by the data processing inequality~\cite{Cover_Thomas_2006} we have that 
\begin{equation}
    I(\mathbf{X};\mathbfcal{Y}_i) \geq I(\mathbf{X};\mathbfcal{Y}_{i+1}),
    \label{eq:dpi}
\end{equation}
where $I(\cdot \, ; \, \cdot)$ is the mutual information. Thus, deeper layers contain less information about the input and are therefore more resistant to \acp{MIA}. Furthermore, as proved in~\cite{scalable_coding_Hyomin}, deeper layers are more compressible at the same level of model accuracy. As a result, deeper layers are more preferable as split points for collaborative intelligence in terms of both privacy and compression efficiency. On the other hand, splitting a \ac{DNN} model at a deeper layer means that the part of the model on the edge device gets larger, increasing its computational requirements. Hence, the actual choice of the split point is a balance between compression and privacy benefits on the one hand, and computational cost on the other. 

\subsection{Compression}
The goal of compression is to minimize the number of bits (bitrate) used to encode the data while preserving the utility of the decoded data. In \emph{lossless} compression, input data can be reconstructed exactly. In \emph{lossy} compression, which is able to achieve higher compression ratios, an exact reconstruction of the input data is not possible. Hence, in addition to minimizing the rate, the goal of lossy compression is to preserve data utility.

In conventional lossy compression, data utility is measured by the deviation between the input data $X$ and the reconstructed data $\widehat{X}$ via a distortion function $d(X,\widehat{X})$.
Hence, conventional compression is usually accomplished by minimizing the following Lagrangian cost:
\begin{equation}
    \mathcal{L}(\lambda) = R + \lambda \cdot \mathbb{E} \left[d(X,\widehat{X})\right],
    \label{eq:RD_lagrange}
\end{equation}
where $R$ is the rate, $\mathbb{E}[\cdot]$ is the expectation operator, and $\lambda$ is the Lagrange multiplier that controls the trade-off between rate and distortion. Applying this to the collaborative intelligence framework where features $\mathbfcal{Y}$ are encoded gives 
\begin{equation}
    \mathcal{L}(\lambda) = R + \lambda \cdot \mathbb{E} \left[d(\mathbfcal{Y},\widehat{\mathbfcal{Y}})\right].
    \label{eq:RD_lagrange_features}
\end{equation}

However, in \emph{coding for machines}, data utility is related to how accurately the machine analysis task can be accomplished based on decoded data, rather than how accurately the input data (or features) can be reconstructed. In fact, it was shown in~\cite{Alon_RD_theor} that computing the feature distortion in deeper layers yields better rate-accuracy trade-off, as deeper layers have less task-irrelevant information. Ultimately, the best performance is achieved when task error at the output of the back-end is used as the distortion. Hence, if $\widehat{\mathbf{V}}=f_2(\widehat{\mathbfcal{Y}};\theta_2)$ is the back-end output based on decoded features $\widehat{\mathbfcal{Y}}$ and $d_V(\mathbf{V},\widehat{\mathbf{V}})$ is the task error, then the Lagrangian cost becomes
\begin{equation}
    \mathcal{L}(\lambda) = R + \lambda \cdot \mathbb{E} \left[d_V(\mathbf{V},\widehat{\mathbf{V}})\right].
    \label{eq:RD_lagrange_cfm_gt}
\end{equation}
This Lagrangian cost is the motivation for some of the loss functions we use for training our autoencoder, as discussed in Section~\ref{subsec:loss_func}.

\subsection{Privacy}
\label{subsec:it_form}
Resisting against \ac{MIA} is one of our main objectives in this paper. This involves increasing uncertainty in the input when the feature tensor at the split point is known. In the information theory terminology, this can be equivalently expressed as maximizing the conditional entropy $H(\mathbf{X}|\mathbfcal{Y})$. At the same time, it is crucial to maintain the utility of the \ac{DNN} model, which involves inferring the precise ground-truth labels $\mathbf{V}$ from $\mathbfcal{Y}$, or alternatively minimizing $H(\mathbf{V}|\mathbfcal{Y})$.

\begin{figure}[!t]
    \centering
    \includegraphics[width=0.5\columnwidth]{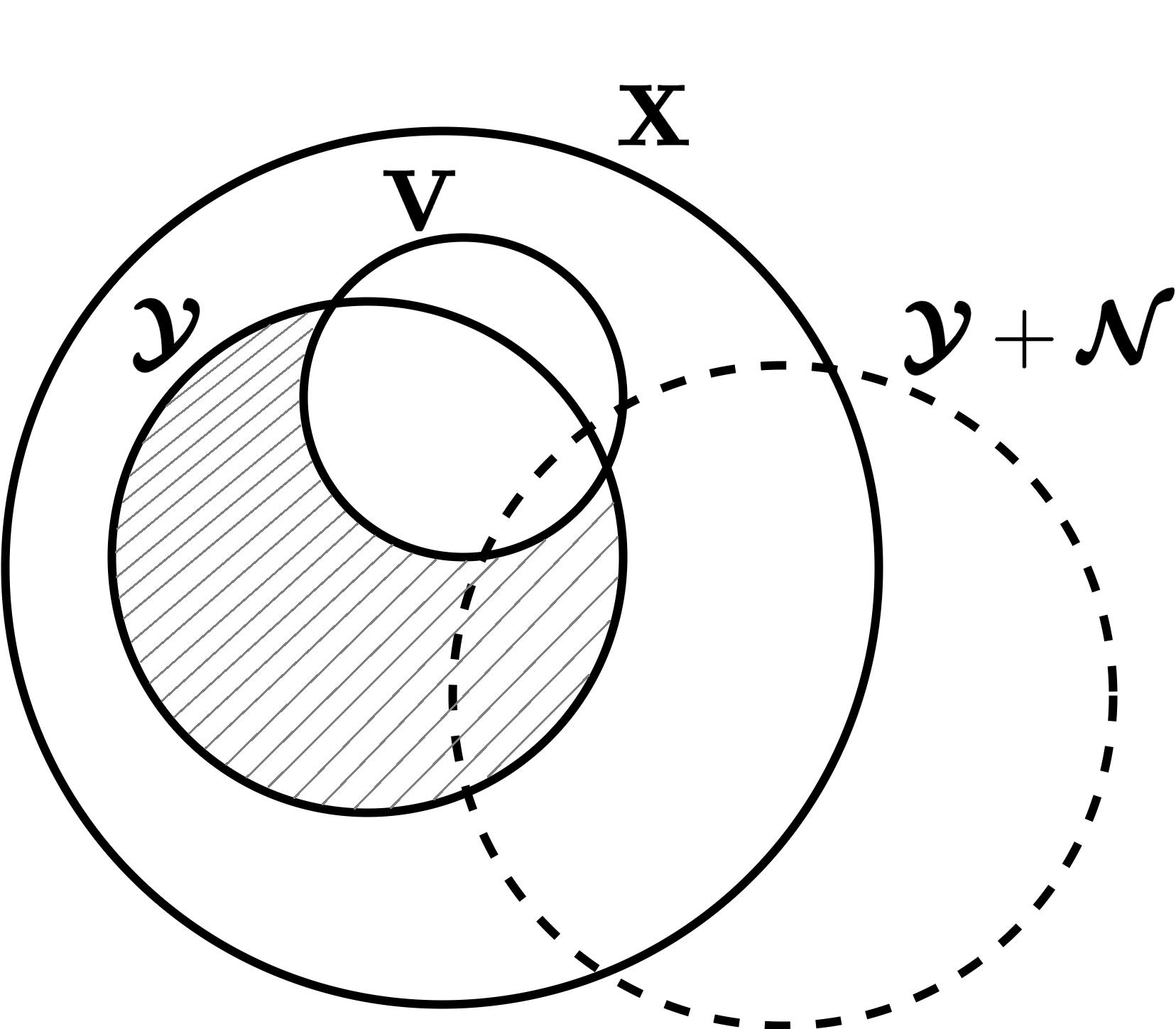}
    \caption{The Venn diagram of the entropy for the involved random vectors.}
    \label{fig:venn}
\end{figure}

The information quantities involved in this problem setting are illustrated in Fig.~\ref{fig:venn}, and refer to the processing chain $\mathbf{X} \to \mathbfcal{Y} \to \overline{\mathbf{V}}$, where $\mathbf{X}$ is the input image, $\mathbfcal{Y}$ is the latent representation (feature tensor) at the split point, and $\overline{\mathbf{V}}$ are the inferred labels, which are ideally close to the ground-truth labels $\mathbf{V}$ (i.e., $\overline{\mathbf{V}}\approx \mathbf{V}$). Note that there is no uncertainty about $\mathbfcal{Y}$ when $\mathbf{X}$ is given, since the trained \ac{DNN} has fixed weights. Hence, $H(\mathbfcal{Y}|\mathbf{X})=0$, which means that the information contained in $\mathbfcal{Y}$ is a subset of the information in $\mathbf{X}$, so $\mathbfcal{Y}$ is shown as a subset of $\mathbf{X}$ in Fig.~\ref{fig:venn}. Moreover, there is no uncertainty in the ground-truth labels when the input is known; thus $H(\mathbf{V}|\mathbf{X})=0$, and $\mathbf{V}$ is shown as a subset of $\mathbf{X}$. Note that inferred labels $\overline{\mathbf{V}}$ (not shown in the figure, for simplicity) are a subset of $\mathbfcal{Y}$, but the ground-truth labels $\mathbf{V}$ are not, unless the DNN model is always perfectly accurate. Hence, in Fig.~\ref{fig:venn}, $\mathbf{V}$ is shown with some overlap with $\mathbfcal{Y}$, but not as a subset of $\mathbfcal{Y}$. Again, consider a trained feedforward \ac{DNN} with fixed parameters (weights) $\theta$. We show that the following lemma holds.

\begin{lemma}
\label{lemma_1}
    $H(\mathbfcal{Y}|\mathbf{V}) = H(\mathbf{V}|\mathbfcal{Y}) - H(\mathbf{X}|\mathbfcal{Y}) + H(\mathbf{X}|\mathbf{V}).$
\end{lemma}

\begin{proof} By the chain rule for mutual information~\cite{Cover_Thomas_2006}
\[I(\mathbfcal{Y}, \mathbf{V}; \mathbf{X}) = I(\mathbfcal{Y};\mathbf{X}) + I(\mathbf{V};\mathbf{X}|\mathbfcal{Y})\]
\[I(\mathbf{V}, \mathbfcal{Y} ; \mathbf{X}) = I(\mathbf{V};\mathbf{X}) + I(\mathbfcal{Y};\mathbf{X}|\mathbf{V})\]
Thus,
\[I(\mathbfcal{Y};\mathbf{X}) + I(\mathbf{V};\mathbf{X}|\mathbfcal{Y}) = I(\mathbf{V};\mathbf{X}) + I(\mathbfcal{Y};\mathbf{X}|\mathbf{V}).\]
Expressing the mutual information terms in terms of entropies, we have
\begin{equation}
\begin{split}
&H(\mathbf{X}) - H(\mathbf{X}|\mathbfcal{Y}) + H(\mathbf{V}|\mathbfcal{Y}) - H(\mathbf{V}|\mathbf{X},\mathbfcal{Y}) = \\ &H(\mathbf{X}) - H(\mathbf{X}|\mathbf{V}) + H(\mathbfcal{Y}|\mathbf{V}) - H(\mathbfcal{Y}|\mathbf{X},\mathbf{V}). 
\label{eq:MI_expansion_lemma_1}
\end{split}
\end{equation}
As mentioned earlier, $H(\mathbfcal{Y}|\mathbf{X})=0$ and $H(\mathbf{V}|\mathbf{X})=0$; thus $H(\mathbfcal{Y}|\mathbf{X},\mathbf{V})=0$ and $H(\mathbf{V}|\mathbf{X},\mathbfcal{Y})=0$. Using these in~(\ref{eq:MI_expansion_lemma_1}) and canceling the common $H(\mathbf{X})$ term gives
\[H(\mathbf{V}|\mathbfcal{Y}) - H(\mathbf{X}|\mathbfcal{Y}) = H(\mathbfcal{Y}|\mathbf{V}) - H(\mathbf{X}|\mathbf{V})\]
from which the claimed equality follows by rearrangement.
\end{proof}

Note that $H(\mathbf{X}|\mathbf{V})$, which is the uncertainty in the input given a ground-truth label, is a function of the data(set) distribution and is independent of the \ac{DNN} model itself. Therefore, it is a constant that cannot be changed through \ac{DNN} construction and training. Then, according to Lemma~\ref{lemma_1}, maximizing $H(\mathbf{X}|\mathbfcal{Y})$ and minimizing $H(\mathbf{V}|\mathbfcal{Y})$ implies that $H(\mathbfcal{Y}|\mathbf{V})$ becomes as small as possible.
In other words, in the optimal scenario, only the information relevant to the ground-truth should be preserved within $\mathbfcal{Y}$, while irrelevant information (the dashed area in Fig.~\ref{fig:venn}) should be removed. According to the results in~\cite{scalable_coding_Hyomin}, the irrelevant information indeed reduces from layer to layer, which lends further support to the conclusion that deeper layers are more resilient against \ac{MIA}.

A related concept is the information bottleneck~\cite{information_bottleneck}, which is formulated as a minimization problem
\begin{equation}
    \min_{p(\mathbfcal{Y}|\mathbf{X}):~ I(\mathbf{V}; \mathbfcal{Y}) \ge C} I(\mathbf{X}; \mathbfcal{Y}),
\label{eq:inf_bott}
\end{equation}
where $p(\mathbfcal{Y}|\mathbf{X})$ is the conditional distribution of $\mathbfcal{Y}$ given $\mathbf{X}$ and corresponds to the front-end $f_1(\cdot \, ; \theta_1)$ part of the network in our setting, while $C$ is a constant ensuring the minimum (mutual) information retained in $\mathbfcal{Y}$ about the desired output. Since $I(\mathbf{X}; \mathbfcal{Y})=H(\mathbf{X})-H(\mathbf{X}|\mathbfcal{Y})$, $I(\mathbf{V}; \mathbfcal{Y})=H(\mathbf{V})-H(\mathbf{V}|\mathbfcal{Y})$, and $H(\mathbf{X})$ and $H(\mathbf{V})$ are (data-dependent) constants,~(\ref{eq:inf_bott}) can also be rewritten as
\begin{equation}
\label{eq:mod_inf_bott}
    \max_{P(\mathbfcal{Y}|\mathbf{X}):~ H(\mathbf{V}|\mathbfcal{Y}) \le C'} H(\mathbf{X}|\mathbfcal{Y}),
\end{equation}
where $C'$ is a new constant. Hence, information bottleneck -- maximizing $H(\mathbf{X}|\mathbfcal{Y})$ while keeping $H(\mathbf{V}|\mathbfcal{Y})$ small -- is aligned with the privacy-utility discussion related to Lemma~\ref{lemma_1} above.
In our proposed method (Section~\ref{sec:method}), we follow this principle, but use loss terms that are more easily computable than mutual information or conditional entropy.

\subsection{\ac{MIA} Methodologies}
\ac{MIA} is the act of recovering the input to a \ac{DNN} model from the intercepted features. In a collaborative intelligence framework, the adversary may access the intermediate features $y$ sent from the edge device to the cloud. 
\acp{MIA} can be categorized into three classes based on the adversary's level of access to the front-end model $f_1(\cdot \, ; \theta_1)$ and its parameters~\cite{mia_2019}:

\begin{enumerate}
    \item \textbf{White-box attack:} The entire front-end model $f_1(\cdot \, ; \theta_1)$ and its parameters $\theta_1$ are known to the adversary. The attack strategy introduced in \cite{mia_2019} involves solving an optimization problem, referred to as regularized Maximum Likelihood Estimation (rMLE).
    \item \textbf{Black-box attack:} The adversary lacks information about $f_1(\cdot \, ; \theta_1)$ but can query the front-end model and access an unlimited number of arbitrary $(x, y)$ pairs. In this scenario, an Inverse Network $f_1^{-1}(\cdot \, ; \theta_1')$ is trained by the adversary using the ground-truth pairs of $(x, y)$, aiming to achieve $f_1^{-1}(y; \theta_1') \approx x$.
    \item \textbf{Query-free attack:} The adversary neither knows $f_1(\cdot \, ; \theta_1)$ nor can query it. However, it has full access to the back-end model. In this setting, a shadow model $g_1(\cdot \, ; \omega_1)$ is first constructed such that $f_2(g_1(x; \omega_1); \theta_2) \approx f_2(y; \theta_2) = f_2(f_1(x; \theta_1); \theta_2)$. This enables $g_1(\cdot \, ; \omega_1)$ to function as a substitute for $f_1(\cdot \, ; \theta_1)$, which can be used to run rMLE or train the Inverse Network.
\end{enumerate}

The white-box setting is the most favorable from the adversary's point of view. However, the experiments in~\cite{mia_2019} indicate that the effectiveness of the black-box attack using the Inverse Network is almost the same as that of the white-box attack using rMLE, and even better when the Inverse Network is trained on the same dataset as the original model. In this paper, we aim to counter the most powerful of these attacks, which is accomplished using the Inverse Network technique trained on the original dataset, in a white-box setting, where the adversary knows some information about the front-end model architecture.

\subsection{Model Inversion Defense Methodologies}
As previously mentioned, in the literature, \ac{DP} and noise obfuscation are the most common solutions to defend against \ac{MIA}~\cite{diff_privacy_iot, not_just_privacy, pricure}. However, adding independent noise to the intermediate features raises uncertainty in both $\mathbf{X}$ and $\mathbf{V}$. 
To see this, consider the following lemma.

\begin{lemma}
\label{lemma_noise}
    Let $\mathbfcal{Y}$, $\mathbf{Z}$ and $\mathbfcal{N}$ be random vectors with $\mathbfcal{N}$ having the same dimension as $\mathbfcal{Y}$ and independent of both $\mathbfcal{Y}$ and $\mathbf{Z}$. Then 
    \begin{equation}
        H(\mathbf{Z}|\mathbfcal{Y} + \mathbfcal{N}) \geq  H(\mathbf{Z}|\mathbfcal{Y}).
    \label{eq:noise_effect}
    \end{equation} 
\end{lemma}

\begin{proof}
Since conditioning does not increase entropy, we have that $H(\mathbf{Z}|\mathbfcal{Y} + \mathbfcal{N}) \geq H(\mathbf{Z}|\mathbfcal{Y} + \mathbfcal{N}, \mathbfcal{N})$. Now that $\mathbfcal{N}$ is one of the conditions, it can be subtracted from the other condition without changing the conditional entropy:  
\[H(\mathbf{Z}|\mathbfcal{Y} + \mathbfcal{N}, \mathbfcal{N}) = H(\mathbf{Z}|(\mathbfcal{Y} + \mathbfcal{N}) - \mathbfcal{N}, \mathbfcal{N}) = H(\mathbf{Z}|\mathbfcal{Y}, \mathbfcal{N}).\]
Finally, since $\mathbfcal{N}$ is independent of $\mathbf{Z}$, we have that $H(\mathbf{Z}|\mathbfcal{Y}, \mathbfcal{N}) = H(\mathbf{Z}|\mathbfcal{Y})$, which completes the proof that $H(\mathbf{Z}|\mathbfcal{Y} + \mathbfcal{N}) \geq H(\mathbf{Z}|\mathbfcal{Y})$.
\end{proof}

Replacing $\mathbf{Z}$ by either $\mathbf{X}$ or $\mathbf{V}$ in Lemma~\ref{lemma_noise} shows that noise obfuscation of features increases both $H(\mathbf{X}|\mathbfcal{Y})$ and $H(\mathbf{V}|\mathbfcal{Y})$, based on which a rough representation of $(\mathbfcal{Y} + \mathbfcal{N})$ entropy is depicted on Fig.~\ref{fig:venn}. That is to say, \ac{MIA} becomes more difficult (higher $H(\mathbf{X}|\mathbfcal{Y})$), but the model becomes less accurate (higher $H(\mathbf{V}|\mathbfcal{Y})$). Ideally, we want to maximize $H(\mathbf{X}|\mathbfcal{Y})$ while minimizing $H(\mathbf{V}|\mathbfcal{Y})$, as discussed in Section~\ref{subsec:it_form}. For this reason, our proposed method is not based on noise obfuscation, but adversarial training with suitably chosen loss functions that promote a high value of $H(\mathbf{X}|\mathbfcal{Y})$ and a low value of $H(\mathbf{V}|\mathbfcal{Y})$.

\section{Proposed Methods}
\label{sec:method}

We chose object detection by YOLOv5~\cite{glenn_jocher_2022_6222936} as the machine vision task, although the overall collaborative intelligence framework is also applicable to other \ac{DNN} models and tasks.
There are different versions of YOLOv5 with varying levels of complexity and accuracy,\footnote{https://github.com/ultralytics/yolov5} and we opted for the YOLOv5m model as the basis for our experiments. In our proposed framework, we follow two key objectives: resistance to \ac{MIA} and improving compression efficiency. To achieve these goals, we train an autoencoder, inserted into the the machine vision model, together with an auxiliary network, called Reconstruction Network (RecNet), aimed at reconstructing the input, in an adversarial manner~\cite{GAN}.

\begin{figure*}[!t]
    \centering
    \includegraphics[width=\textwidth]{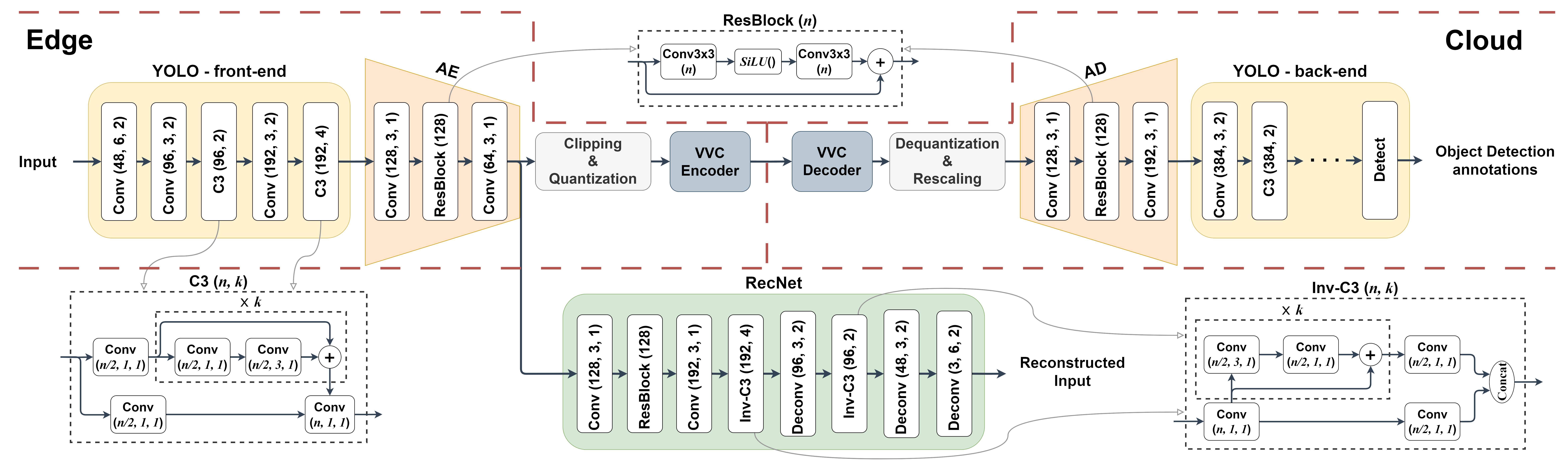}
    \caption{The overall block diagram of the proposed method. $Conv(n,k,s)$ is a 2D Convolution layer followed by a Batch normalization layer and a $SiLU()$ activation, with $n$ being the number of output channels, a kernel size of $k\times k$, and stride=$s$. $Deconv(n,k,s)$ is the same as $Conv(n,k,s)$ except with a Convolution Transpose layer. Note that the $Conv$ layers in the autoencoder do not have batch normalization, and the last $Conv$ in AE does not have the $SiLU()$ activation either. $Conv3\times 3(n)$ is a single $3\times 3$ Convolution layer with stride=1 and $n$ output channels. ``$\times k$" indicates $k$ times repetition of the blocks specified by a dashed frame.}
    \label{fig:block_diagram}
\end{figure*}

The overall block diagram of our design is shown in Fig.~\ref{fig:block_diagram}. 
It is worth mentioning that the autoencoder is trained offline prior to deployment. At the inference stage, the YOLO front-end and the encoder portion of the autoencoder are deployed on the edge device, along with the VVC encoder, while the VVC decoder, the decoder portion of the autoencoder, and the YOLO back-end are deployed in the cloud. ``RecNet'' is used only during training to improve the adversarial robustness of the autoencoder, and is not part of the deployed pipeline. These components are further elaborated on below.

\subsection{Split Point}
For collaborative edge-cloud deployment, the YOLOv5m model should be split into two parts: the front-end, which is to be deployed on the edge device, and the back-end, which would reside in the cloud. As discussed in Section~\ref{subsec:ci_form}, a deeper split point is more desirable in terms of both compression efficiency and privacy. However, limited computing power and energy considerations of edge devices favor choosing a shallower split point. Moreover, the YOLOv5m model branches out at layer 5. Picking a split point deeper than layer 5 would require encoding and transmitting multiple sets of feature tensors, which would increase complexity and make compression more challenging. Thus, we opted to split the model at layer 5.

\subsection{Autoencoder}
At the split point, an autoencoder is integrated and trained with the goal of reducing the dimensionality of the original latent space and generating bottleneck features with improved compressibility and resilience to \ac{MIA}. This is a plug-and-play strategy and can be readily applied to other models and computer vision tasks without retraining the original vision model. The autoencoder's architecture is shown in Fig.~\ref{fig:block_diagram}: its encoder portion is denoted as AE, and the decoder is referred to as AD. 

The role of AE is to transform the YOLO's native latent space at the split point (with $192$ channels) into lower-dimensional bottleneck features (here, with $64$ channels). The spatial dimensions remain unchanged to maintain spatial precision, which is crucial for subsequent object detection. The resulting bottleneck feature tensor is tiled and pre-quantized to 8 bits per element, similarly to~\cite{Hyomin_deep_feature_compression}, and then encoded using \ac{VVC}-Intra~\cite{VVC_overview}. On the cloud side, the encoded bitstream is decoded by a VVC decoder and AD, and then fed to the YOLOv5m back-end for object detection.

\subsection{Loss functions}
\label{subsec:loss_func}
To train the autoencoder, we employ three main loss functions. Each of these loss functions is meant to encourage a certain aspect of the bottleneck features, as explained below.

\begin{itemize}
    \item \textbf{Object detection:} This is the native object detection loss function used in YOLOv5~\cite{glenn_jocher_2022_6222936}:
    \begin{equation}
        L_{obj\_det} = w_{obj}\cdot L_{obj} + w_{box}\cdot L_{box} + w_{cls}\cdot L_{cls},
        \label{eq:obj_det}
    \end{equation}
    where $L_{obj}$, $L_{box}$, and $L_{cls}$ are objectness, bounding box localization, and classification losses that are combined with corresponding balancing weights. This loss term is the task error $\mathbb{E} \left[d_V(\mathbf{V},\widehat{\mathbf{V}})\right]$ from~(\ref{eq:RD_lagrange_cfm_gt}). Improving object detection is related to the maximization of the $I(\mathbf{V}; \mathbfcal{Y})$ term in~(\ref{eq:inf_bott}), since accurate detection requires a sufficiently high mutual information between $\mathbfcal{Y}$ and $\mathbf{V}$.
    
    \item \textbf{Compressibility:} $L_{cmprs}$ is a loss function borrowed from~\cite{compressible_loss} and aims to enhance the compressibility of feature tensors using traditional image codecs:
    \begin{equation}
        L_{cmprs}=\frac{1}{H\cdot W\cdot C}\sum_{i=1}^{C}\|\mathbf{T}_i \|_1.
        \label{eq:compressibility_loss}
    \end{equation}
    Here, $H$, $W$, and $C$ are the height, width, and number of bottleneck feature channels, respectively;  $\mathbf{T}_i=\texttt{DCT}\{\mathbf{Z}_i\}$, where $\texttt{DCT}$ is the Discrete Cosine Transform and $\mathbf{Z}_i$ contains the residuals of horizontal and vertical prediction within each feature channel~\cite{compressible_loss}. The purpose of such a loss term is to mimic or approximate the processing steps performed by a typical conventional image encoder in terms of differentiable operations and compute a quantity ($\ell_1$ norm of transform coefficients) that is known to be related to the resulting bitrate. Hence, $L_{cmprs}$ is a proxy for the rate $R$ in~(\ref{eq:RD_lagrange_cfm_gt}). Since we use \ac{VVC}-Intra~\cite{VVC_overview} to code bottleneck features, such a loss term is appropriate for our design. Note that improving compressibility is related to the minimization of the $I(\mathbf{X}; \mathbfcal{Y})$ term in~(\ref{eq:inf_bott}), which is the optimal rate at which $\mathbf{X}$ can be encoded into $\mathbfcal{Y}$~\cite{Cover_Thomas_2006}.
    
    \item \textbf{Reconstruction:} The reconstruction loss function $L_{rec}$ is intended to capture the quality of reconstruction of the input image from the bottleneck features. $L_{rec}$ consists of two terms: the $\ell_1$-norm of the reconstruction error and an edge-centric loss.
    The latter is related to fine details in the image, which are usually associated with private information (e.g., facial features).
    $L_{rec}$ is computed as follows:
    \begin{equation}
        \begin{split}
             L_{rec} = \frac{1}{n}\|{x}-\hat{x}\|_1 & +  \frac{\beta}{n}\|S_h*{x}-S_h*\hat{x}\|_1 \\ 
             & + \frac{\beta}{n}\|S_v*{x}-S_v*\hat{x}\|_1,
        \end{split}
        \label{eq:L_rec}
     \end{equation}
     where $x$ and $\hat{x}$ are the original and reconstructed image, respectively, $S_h$ and $S_v$ are the horizontal and vertical Sobel filter~\cite{Gonzalez-Woods_2018}, respectively, $*$ is the convolution operator, and $n$ is the total number of tensor elements in the training batch. The value of $\beta$ is empirically set to $5$.
\end{itemize}
    
\subsection{Adversarial Training}
\label{subsec:adv_train}
The training is performed offline in a centralized manner. We have adopted an adversarial training strategy to generate compressible bottleneck features that are resistant to \ac{MIA} while being able to support the object detection task. This is done using the auxiliary RecNet model, the purpose of which is to mimic an adversary performing \ac{MIA} and try to reconstruct the input image from the bottleneck features. As illustrated in Fig.~\ref{fig:block_diagram}, RecNet's architecture roughly mirrors that of the YOLOv5m front-end and AE. This is in line with a white-box attack, where the adversary knows the architecture of the model under attack.

The autoencoder and RecNet are adversarially trained together, with RecNet serving as the ``discriminator'' and YOLO front-end and AE as the feature generator.
Throughout the training process, RecNet tries to recover the input image from the bottleneck features as best it can, while the AE attempts to disrupt RecNet's performance by manipulating the generated bottleneck features. Meanwhile, both AE and AD work to maintain high object detection accuracy. Furthermore, bottleneck features are progressively optimized for compression efficiency via the compressibility loss term during training. The overall loss function is
\begin{equation}
    L_{tot} = L_{obj\_det} + w_{cmprs}\cdot L_{cmprs} - w_{rec}\cdot L_{rec},
    \label{eq:L_tot}
\end{equation}
where $L_{obj\_det}$, $L_{cmprs}$, and $L_{rec}$ are the loss terms defined in Section \ref{subsec:loss_func}. The balancing weights for combining these loss terms are denoted $w_{cmprs}$ and $w_{rec}$. Experiments are performed for different combinations of these weights in Section~\ref{subsec:ablation} to study the effect of each loss term on the overall model performance in terms of privacy and compression efficiency. The impact of the proposed edge-centric loss term in (\ref{eq:L_rec}) is analyzed in Section~\ref{subsec:resistance}.

\begin{figure}[!t]
    \centering
    \subfloat[]{
        \centering
        \includegraphics[width=0.8\columnwidth]{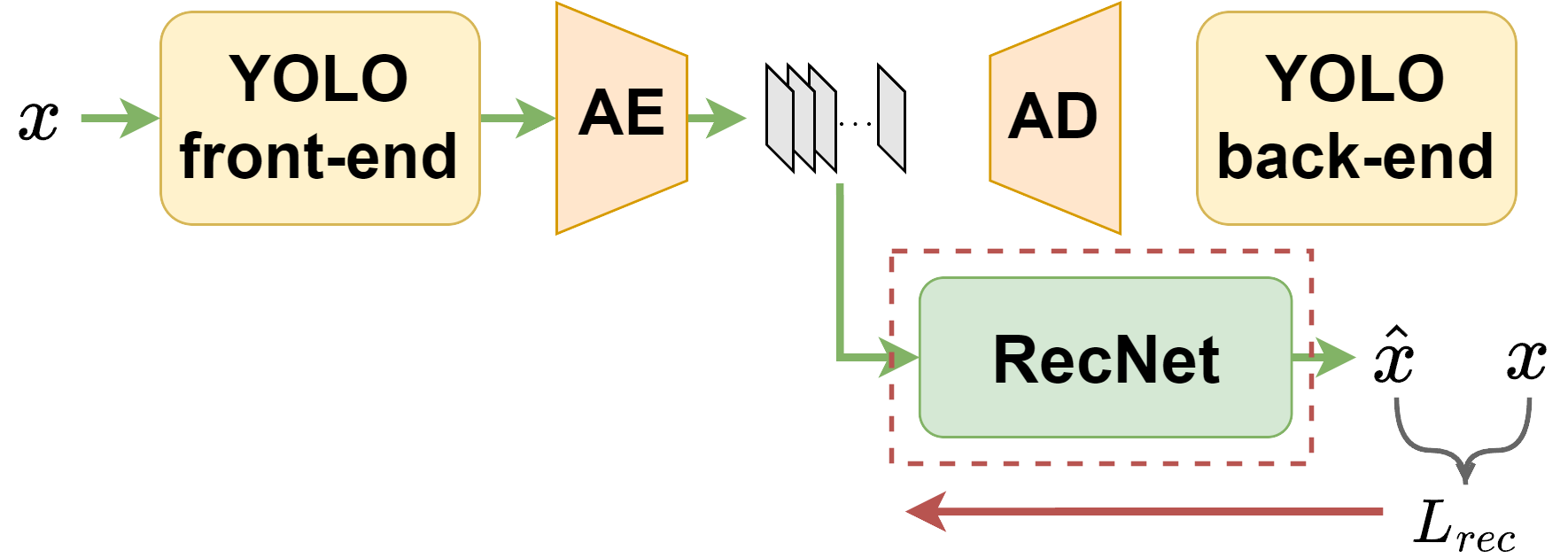}
        \label{subfig:data_flow_a}
    }
    \hfill
    \subfloat[]{
        \centering
        \includegraphics[width=0.93\columnwidth]{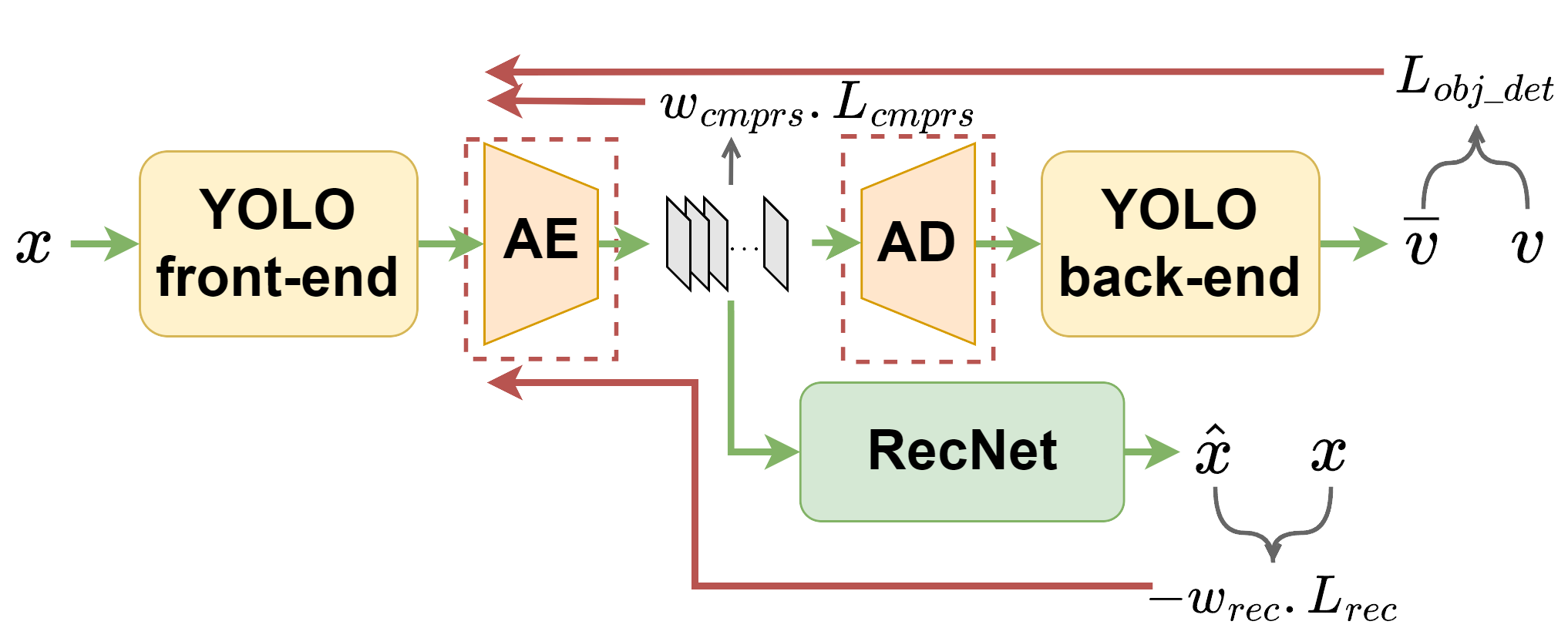}
        \label{subfig:data_flow_b}
    }
    \hfill
    \caption{The proposed adversarial training steps; green and red arrows represent feedforward and backpropagation, respectively. Only the blocks enclosed by the dashed lines are trained at each corresponding step.}
    \label{fig:data_flow}
\end{figure}

The adversarial training takes the form of min-max optimization
\begin{equation}
    \min_{\theta_A} \max_{\theta_R} L_{tot},
    \label{eq:adv_sol}
\end{equation}
where $\theta_A$ are autoencoder's parameters and $\theta_R$ are RecNet's parameters.
Note that the pre-trained YOLOv5m model remains intact, and its weights are frozen throughout the entire training process. Like most adversarial training schemes, the model parameters need to be updated in multiple steps for each data batch, summarized as follows:

\begin{enumerate}
    \item The input images pass through the front-end, AE, and RecNet, and the corresponding reconstruction loss $L_{rec}$ is calculated according to~(\ref{eq:L_rec}) (see Fig.~\ref{subfig:data_flow_a}).
     \item The gradient of $L_{rec}$ backpropagates only through the RecNet and updates its weights. The autoencoder's weights are frozen during this step (see Fig.~\ref{subfig:data_flow_a}).
     \item The same batch of images goes through the whole network, and $L_{tot}$ is computed based on (\ref{eq:L_tot}) (see Fig.~\ref{subfig:data_flow_b}):
     
     \item The gradient of $L_{tot}$ backpropagates through the autoencoder and updates its weights. RecNet's weights remain frozen during this step (see Fig.~\ref{subfig:data_flow_b}). Note that the negative sign of $L_{rec}$ in~(\ref{eq:L_tot}) leads AE to make reconstruction more difficult for RecNet, thereby maximizing $H(\mathbf{X}|\mathbfcal{Y})$, or equivalently, enhancing privacy. Meanwhile, the positive sign of $L_{obj}$ and $L_{cmprs}$ drives AE and AD to improve object detection accuracy and create compressible features, which would result in minimizing $H(\mathbf{V}|\mathbfcal{Y})$ and bitrate, respectively.
\end{enumerate}

\section{Proposed Evaluation Framework}
\label{sec:eval_frame}
\begin{figure}[!t]
    \centering
    \subfloat[]{
        \centering
        \includegraphics[width=0.3\columnwidth]{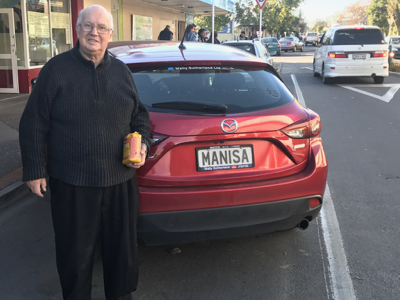}
        \label{subfig:car_face_orig}
    } 
    \hfill
    \subfloat[]{
        \centering
        \includegraphics[width=0.3\columnwidth]{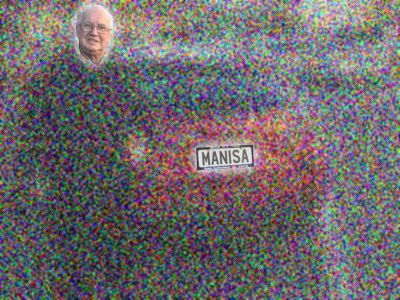}
        \label{subfig:car_face_non_pri}
    } 
    \hfill
    \subfloat[]{
        \centering
        \includegraphics[width=0.3\columnwidth]{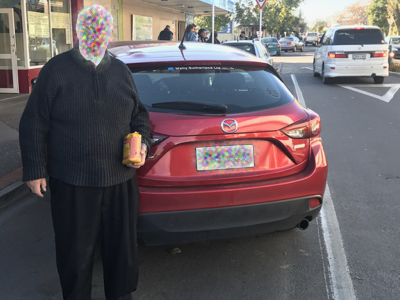}
        \label{subfig:car_face_pri}
    } 
    \caption{Image quality may not be a good proxy for privacy: (a) original image;  (b) low-quality image (PSNR = 28.0~dB) but with private information exposed; (c) higher-quality image (PSNR = 41.4~dB) with private information protected.}
    \label{fig:psnr_privacy}
    \vspace{-0.3cm}
\end{figure}

Measuring the privacy of a coding system has always been a challenge. Many researchers rely on perceptual metrics to quantify how much private data in an image has been preserved. For example, a low value of PSNR or SSIM between the target and the original image is often considered a sign of high privacy protection. However, private data is not uniformly distributed across the entire image, whereas perceptual metrics naturally consider the whole image.
To illustrate why such metrics might not be appropriate proxies for privacy protection, Fig.~\ref{fig:psnr_privacy}(a) shows an image that exposes some private information (face and license plate). In Fig.~\ref{fig:psnr_privacy}(b), noise is added to the entire image except the face and license plate regions, creating low PSNR, but with private information still fully exposed. In Fig.~\ref{fig:psnr_privacy}(c), noise is added only to the face and license plate region, leading to an image with much higher PSNR but with private information now obscured by noise. Therefore, in this example, PSNR is a poor proxy for privacy protection.

For this reason, we propose an evaluation framework in which privacy is measured more directly, instead of relying on proxies such as PSNR and SSIM. Specifically, our evaluation is based on the face and license plate recognition accuracy obtained on the images recovered by running a \ac{MIA} on the bottleneck features. These evaluation metrics are elaborated in Sections~\ref{subsec:face_rec} and \ref{subsec:license_rec}, while Section~\ref{subsec:bench} presents
the benchmark evaluation models.

\subsection{Face Recognition Metric}
\label{subsec:face_rec}
Faces contain critical information through which the identity of a person can be divulged. In order to quantify the privacy level of our proposed method, we measure how accurately a face recognition model can identify faces within the images recovered from bottleneck features by \ac{MIA}. Various face recognition models, especially those based on \acp{DNN}, are available. We employed the implementation of MagFace~\cite{magface}, along with its pre-trained model accessible on GitHub.\footnote{https://github.com/IrvingMeng/MagFace} MagFace is a well-known \ac{DNN}-based face recognition model with high accuracy. As reported in~\cite{magface}, MagFace has achieved $99.83\%$ \textit{verification} accuracy on the LFW dataset~\cite{LFWTech}. 

Face recognition models are often evaluated on the \textit{verification} task, where the goal is to determine whether two distinct images show the same individual. However, privacy primarily concerns the precision with which an individual can be recognized from an image. Hence, we believe \textit{identification} task, as outlined in~\cite{lfw_identification, lfw_identification_2}, is more closely related to the notion of privacy. In the identification task, two sets of images are involved: the gallery set and the probe set. The gallery set comprises sample images of individuals, while the probe set contains images of people of interest. The face recognition model should match the images within the probe set with the corresponding images in the gallery showing the same individuals.

The LFW dataset consists of $13,233$ images belonging to $5,749$ individuals, $1,680$ of which have at least two images available. From these $1,680$ individuals, one of their photos is assigned to the gallery set, and the rest are assigned to the probe set.
Next, we collect the output feature vectors generated by MagFace for all the images and compute the distance between all pairs of members from the gallery and probe sets. The image in the gallery with the lowest feature distance to the currently-investigated probe image is selected as the match. In the end, the identification accuracy is calculated based on the number of correctly detected matches.

\subsection{License Plate Recognition Metric}
\label{subsec:license_rec}
License plate data is another form of sensitive information that can be exploited to track specific vehicles for potentially malicious purposes using citywide traffic monitoring cameras. Hence, license plate recognition accuracy can be considered another direct measure of privacy.
For this purpose, we used LPRNet~\cite{lprnet_license_recognition}, a \ac{DNN}-based license plate recognition model, whose implementation is available on GitHub.\footnote{https://github.com/xuexingyu24/License\_Plate\_Detection\_Pytorch}

In our experiments, the license plate recognition accuracy is calculated on a set of $5,000$ images from the CCPD validation dataset~\cite{ccpd_license_dataset}. To simulate a stronger privacy attack, we fine-tuned the LPRNet model on a set of $5,000$ distorted images from the CCPD training set, which were recovered from the intercepted features. Therefore, all subsequent license plate recognition accuracies are based on the fine-tuned LPRNet.

\subsection{Benchmarks}
\label{subsec:bench}
\begin{figure}[!t]
    \centering
    \subfloat[Benchmark-input]{
        \centering
        \includegraphics[width=0.6\columnwidth]{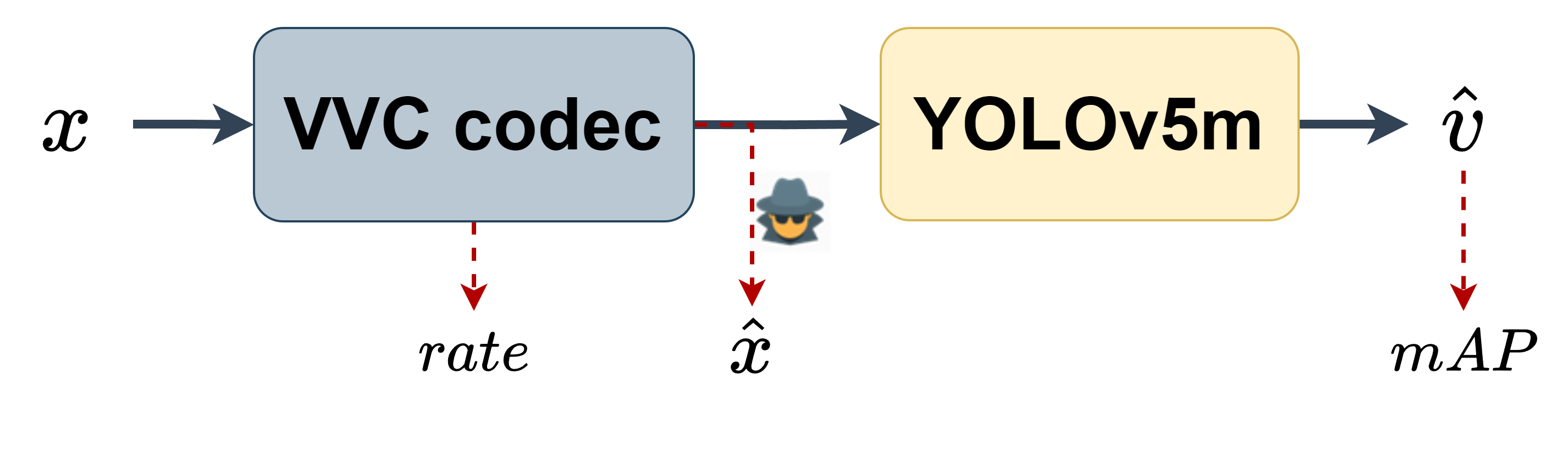}
        \label{subfig:bench_in}
    } 
    \hfill
    \subfloat[Benchmark-latent]{
        \includegraphics[width=0.65\columnwidth]{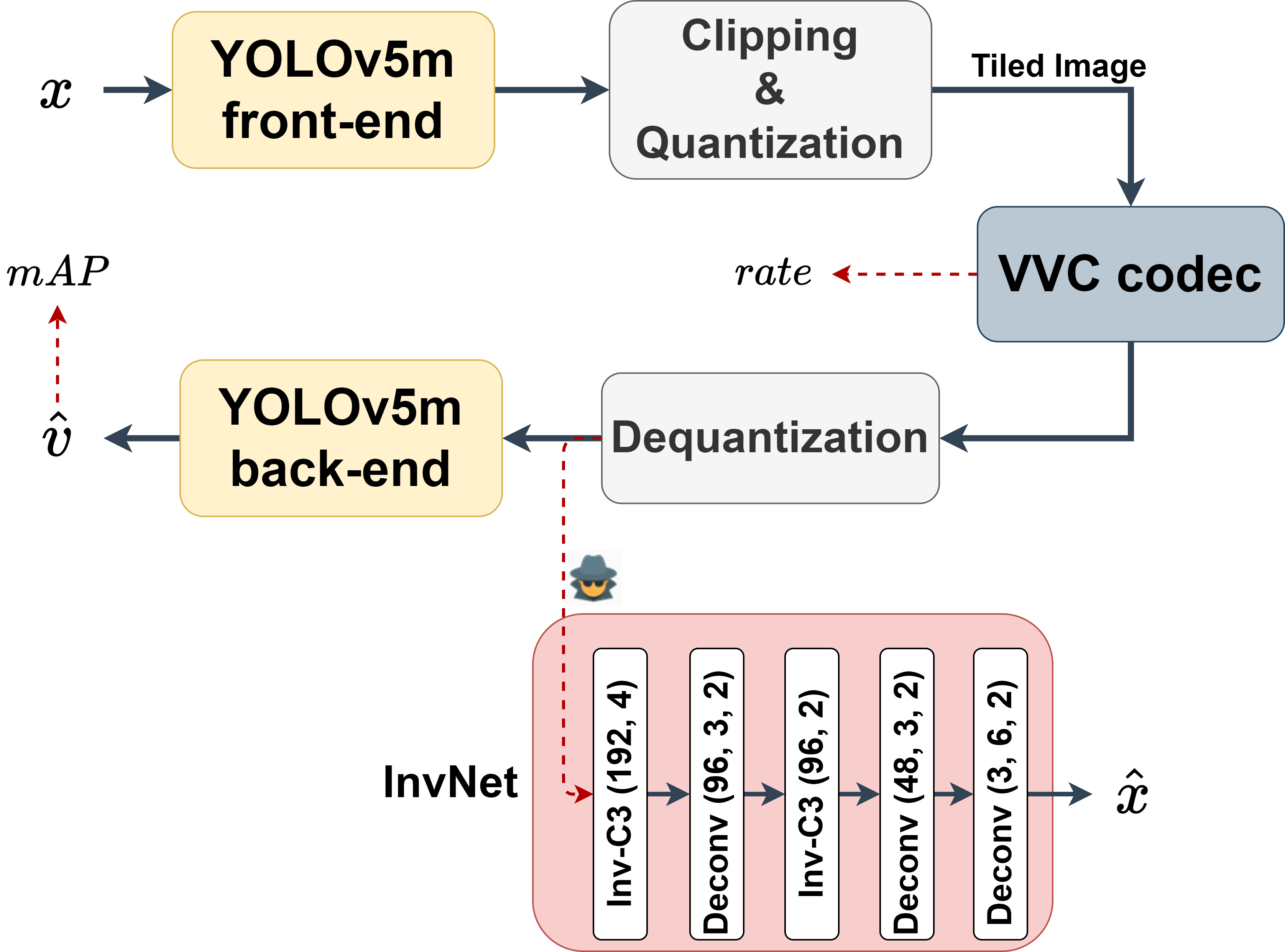}
        \label{subfig:bench_late}
    } 
    \hfill
    \subfloat[Benchmark-bottleneck / Proposed]{
        \includegraphics[width=0.65\columnwidth]{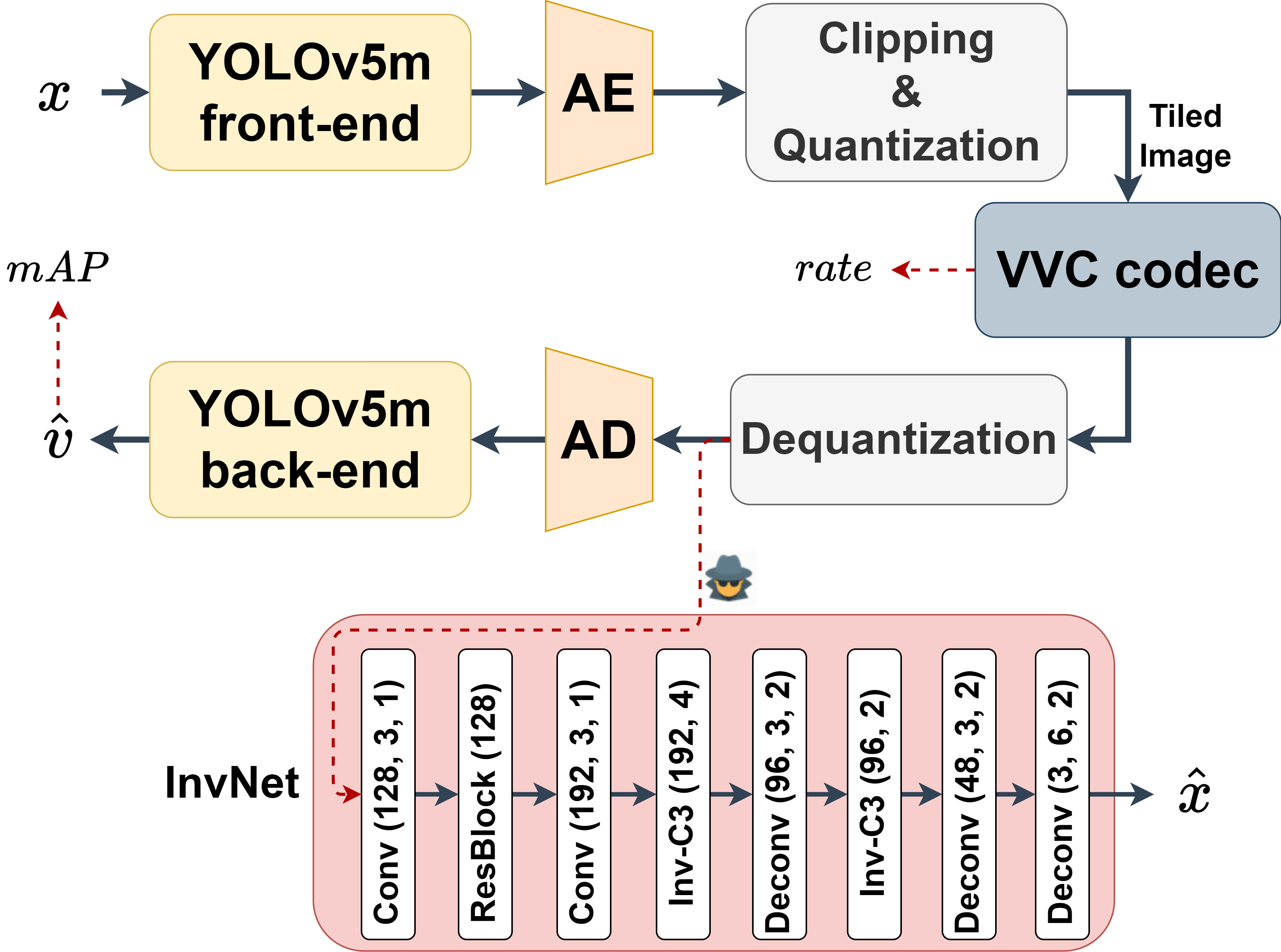}
        \label{subfig:bench_bott}
    }
    \caption{The evaluation pipelines}
    \label{fig:benchmarks}
\end{figure}

We compare the performance of our proposed method with the benchmark pipelines shown in Fig.~\ref{fig:benchmarks}. In ``Benchmark-input,'' the input image is encoded using the VVC codec, and object detection is performed on the decoded image ($\hat{x}$). The accuracy of face and license plate recognition is also measured on the decoded image.
``Benchmark-latent'' deals with the feature channels at YOLOv5's fifth layer, where 192 channels are clipped, quantized to 8 bits, tiled into a 
gray-scale matrix, and then encoded using the VVC codec. The decoded features are then fed to the YOLOv5 back-end for object detection. To measure privacy protection, an approximation to the input image ($\hat{x}$) is created by an inverse network (InvNet), whose architecture mirrors the layers that produce the features, similar to a white-box attack. The accuracy of face and license plate recognition is measured on the recovered image $\hat{x}$.

The pipelines for ``Benchmark-bottleneck'' and our proposed method are shown in Fig.~\ref{fig:benchmarks}(c). In both cases, the 64 bottleneck channels of the autoencoder are tiled into 
grayscale matrix after clipping and quantization, and encoded using the VVC codec. The only difference between the two is that our method's autoencoder is trained adversarially with non-zero weights $w_{cmprs}$ and $w_{rec}$, while the autoencoder in Benchmark-bottleneck is trained only using YOLO's native object detection loss (i.e., $w_{cmprs}=w_{rec}=0$ in~(\ref{eq:L_tot})). The InvNet for the proposed method and ``Benchmark-bottleneck'' mirrors both the YOLOv5m front-end and the autoencoder's encoding part (AE). Again, the accuracy of face and license plate recognition is measured on the recovered image $\hat{x}$.

\section{Experiments}
\label{sec:exp_res}

\begin{figure*}[!t]
    \centering

        \centering
        \includegraphics[width=0.16\textwidth]{figures/3_person_orig.jpg}
    \hfill
        \centering
        \includegraphics[width=0.16\textwidth]{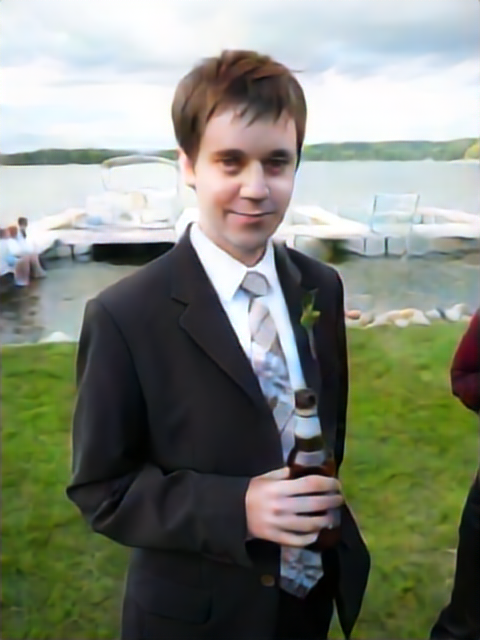}
    \hfill
        \centering
        \includegraphics[width=0.16\textwidth]{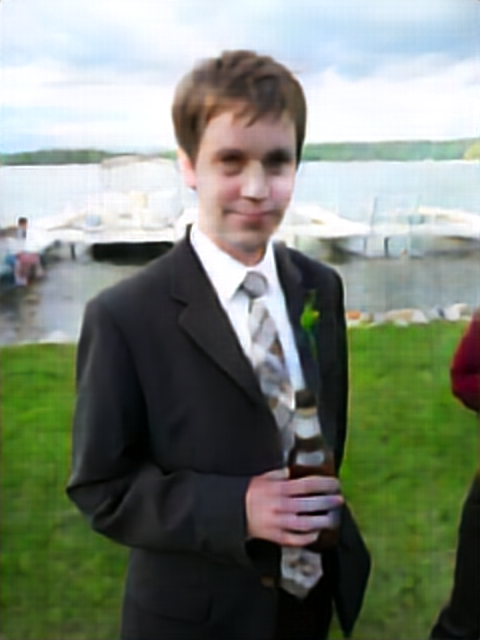}
    \hfill
        \centering
        \includegraphics[width=0.16\textwidth]{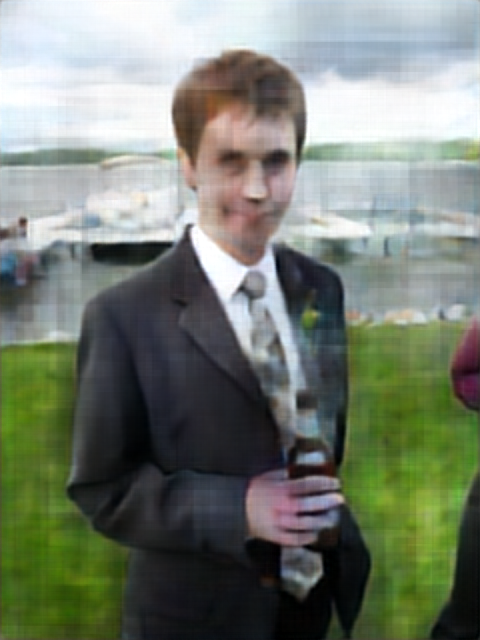}
        \centering
        \includegraphics[width=0.16\textwidth]{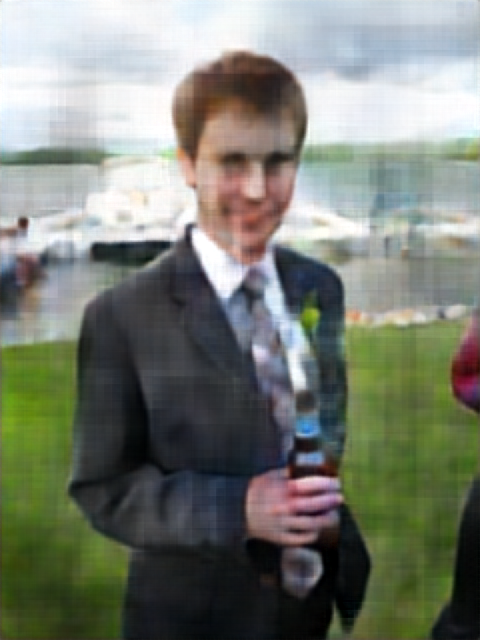}
    \hfill
        \centering
        \includegraphics[width=0.16\textwidth]{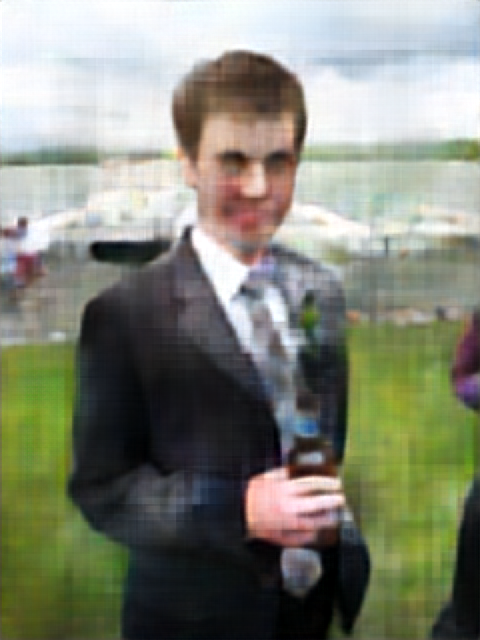}
    \hfill
    

    \subfloat[]{
        \centering
        \includegraphics[width=0.15\textwidth]{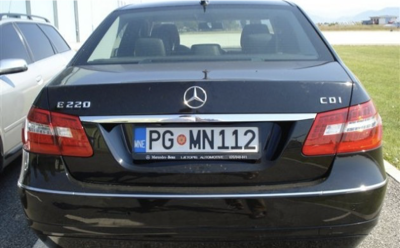}
        \label{subfig:vis-orig}
    }
    \hfill
    \subfloat[]{
        \centering
        \includegraphics[width=0.15\textwidth]{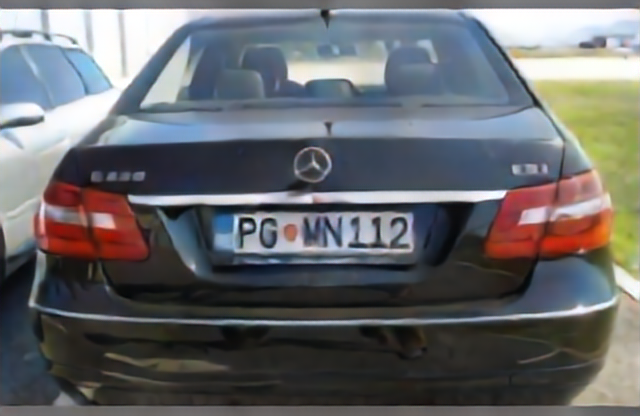}
        \label{subfig:vis-late}
    }
    \hfill
    \subfloat[]{
        \centering
        \includegraphics[width=0.15\textwidth]{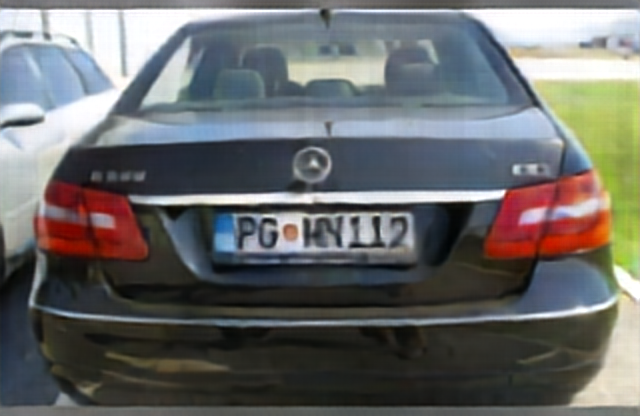}
        \label{subfig:vis-w00}
    }
    \hfill
    \subfloat[]{
        \centering
        \includegraphics[width=0.15\textwidth]{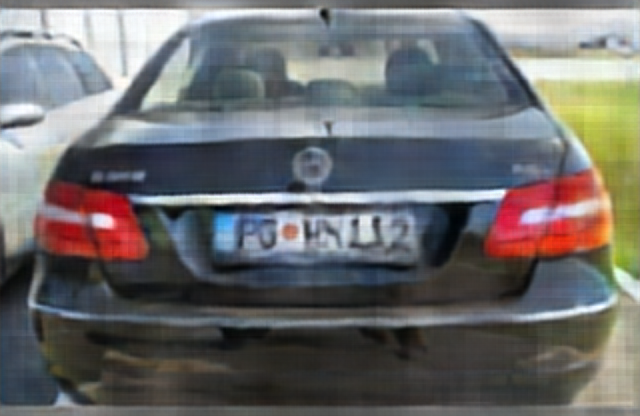}
        \label{subfig:vis-w05}
    }
    \hfill
    \subfloat[]{
        \centering
        \includegraphics[width=0.15\textwidth]{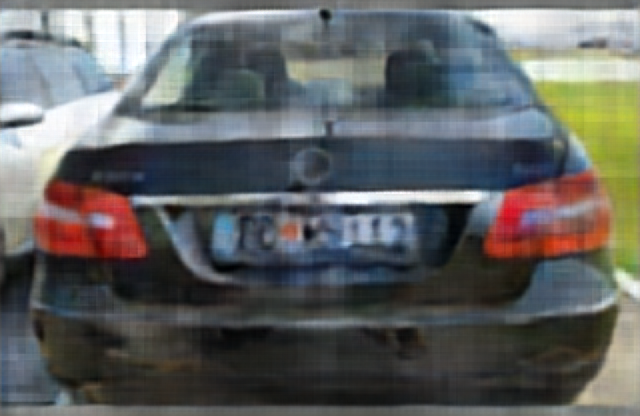}
        \label{subfig:vis-w10}
    }
    \hfill
    \subfloat[]{
        \centering
        \includegraphics[width=0.15\textwidth]{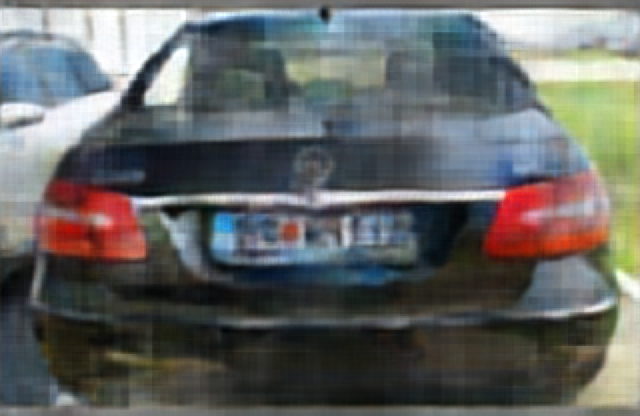}
        \label{subfig:vis-w20}
    }

    \caption{Examples of images recovered through \ac{MIA}: (a) original image  (b) Benchmark-latent (c) Benchmark-bottleneck (d) $(w_{rec},w_{cmprs})=(0.5,0)$ (e) $(w_{rec},w_{cmprs})=(1.0,0)$ (d) $(w_{rec},w_{cmprs})=(2.0,0)$}
    \label{fig:vis-examples}  
    \vspace{-0.3cm}
\end{figure*}

The training was performed on the COCO-2017 object detection dataset \cite{COCO} using an NVIDIA Tesla V100-SXM2 GPU with 32GB of memory. For the proposed method, in the first step, only the autoencoder was trained with $L_{obj\_det}$ for 50 epochs. Next, we trained the RecNet with $L_{rec}$ for 20 epochs, while keeping the autoencoder's weights frozen to those obtained in the first step. Finally, with the autoencoder and RecNet initialized to the weights obtained in the first two steps, the adversarial training was conducted with various values of $w_{cmprs}$ and $w_{rec}$, as detailed in Section~\ref{subsec:adv_train}, for 40 epochs.
In all stages, we used the Stochastic Gradient Descent (SGD) optimizer with an initial learning rate of 0.01, changing with a cosine learning rate decay~\cite{bag_of_tricks_2019_CVPR} over the training epochs. VVenC~\cite{VVenC} was employed as the VVC-Intra codec.

As noted in~\cite{Bob_Collab_Clip}, feature compression performance can be impacted by the selected clipping range. We empirically selected $\pm6\sigma$ as the clipping range, where $\sigma$ is the standard deviation of the to-be-coded feature tensor values on the COCO validation set.

As mentioned earlier, RecNet is an auxiliary DNN model exploited in the adversarial training stage and is not part of the final system for collaborative object detection. In a real-world situation, an adversary would train their own Inverse Network (InvNet) using input-feature pairs generated by querying the front-end model deployed on the edge device. Hence, to mimic this scenario, we also train new, randomly initialized InvNets on the bottleneck features generated by the autoencoders, in order to recover the input. Note that in the following experiments, each autoencoder trained with different hyperparameters has a unique trained InvNet associated with its bottleneck features. These InvNets and the benchmarks' InvNets are trained with an $\ell_1$-norm loss of the error (the first term in~(\ref{eq:L_rec})) on the COCO dataset, the same dataset used for training the autoencoder and YOLOv5 model. The correspondence of datasets for training the main model and InvNets could result in better efficacy of \ac{MIA} according to~\cite{mia_2019}. Therefore, the resistance of the models would be examined against a stronger \ac{MIA}.

\subsection{Resistance to \ac{MIA}}
\label{subsec:resistance}
Experiments in this section focus on the resistance of the models against \ac{MIA}, without involving feature compression by the VVC codec. In other words, the gray-toned blocks within the evaluation pipelines shown in Fig.~\ref{fig:benchmarks} are all bypassed. Fig.~\ref{fig:vis-examples} shows images recovered from bottleneck features by running a \ac{MIA}. These images correspond to the benchmark pipelines and our proposed autoencoder, which was adversarially trained with different values of $w_{rec}$ while setting $w_{cmprs}$ to $0$. 

As seen in Fig.~\ref{fig:vis-examples}, the adversary's Inverse Networks are able to recover the input images from the Benchmark-latent and Benchmark-bottleneck to a good extent. In particular, facial and license plate details are still visible in these images. This suggests that even with a reduction in the number of feature channels from 192 at the YOLOv5m model's split point to 64 at the autoencoder's bottleneck, there remains sufficient information for license plate recognition and person identification. On the other hand, adversarially trained autoencoders, particularly those trained with higher values of $w_{rec}$, are capable of effectively removing this information. This is because higher values of $w_{rec}$ emphasize $L_{rec}$ in the total loss (see~(\ref{eq:L_tot})), which encourages AE to cause a drop in the quality of the recovered images. This degradation in quality is more concentrated near the edges and textured areas, due to the use of Sobel filters in the reconstruction loss as presented in~(\ref{eq:L_rec}).

\begin{figure}[!t]
    \centering
    \subfloat[]{
        \centering
        \includegraphics[width=0.85\columnwidth]{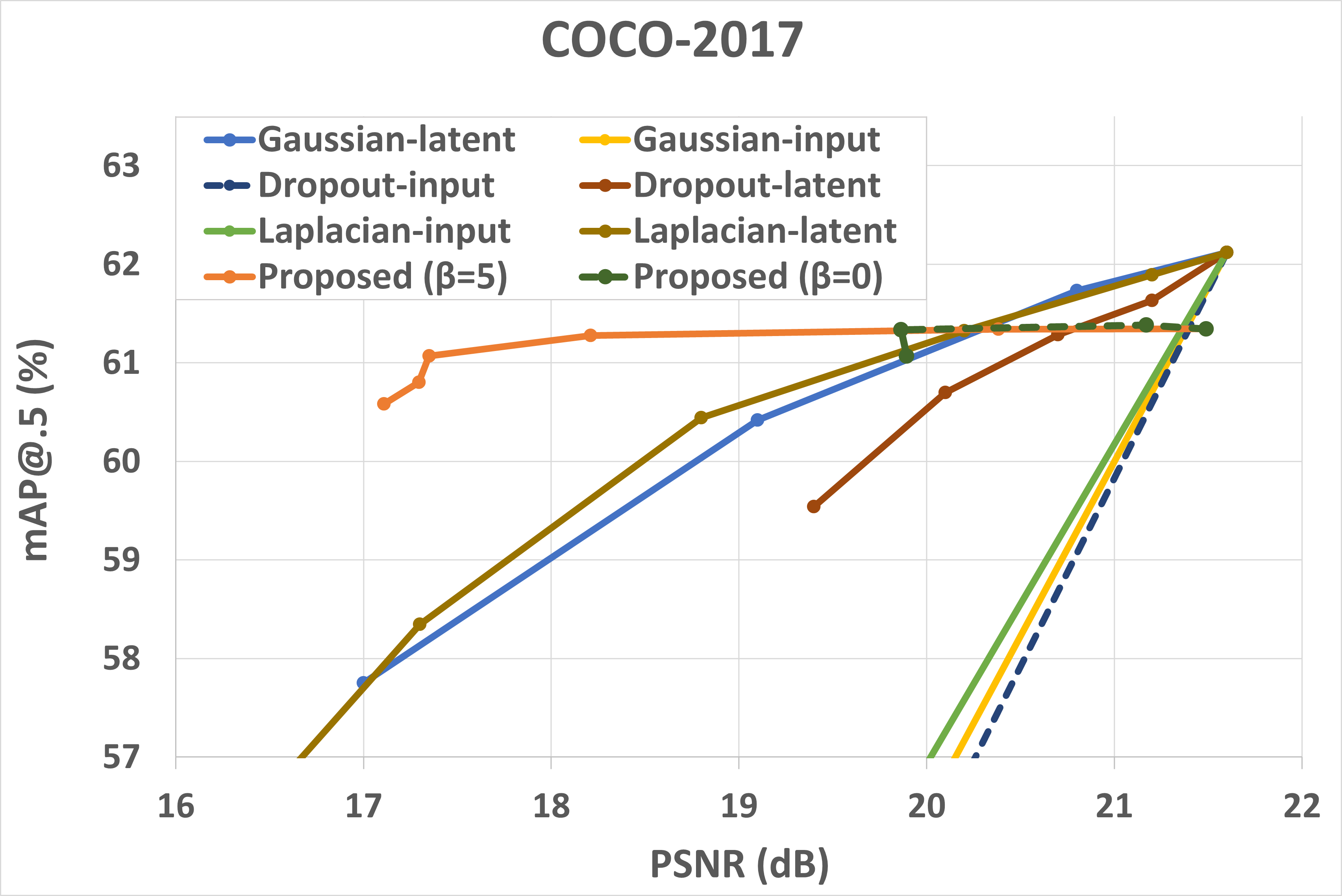}
        \label{subfig:wo_cmprs_psnr-acc}
    } 
    \hfill
    \subfloat[]{
        \centering
        \includegraphics[width=0.85\columnwidth]{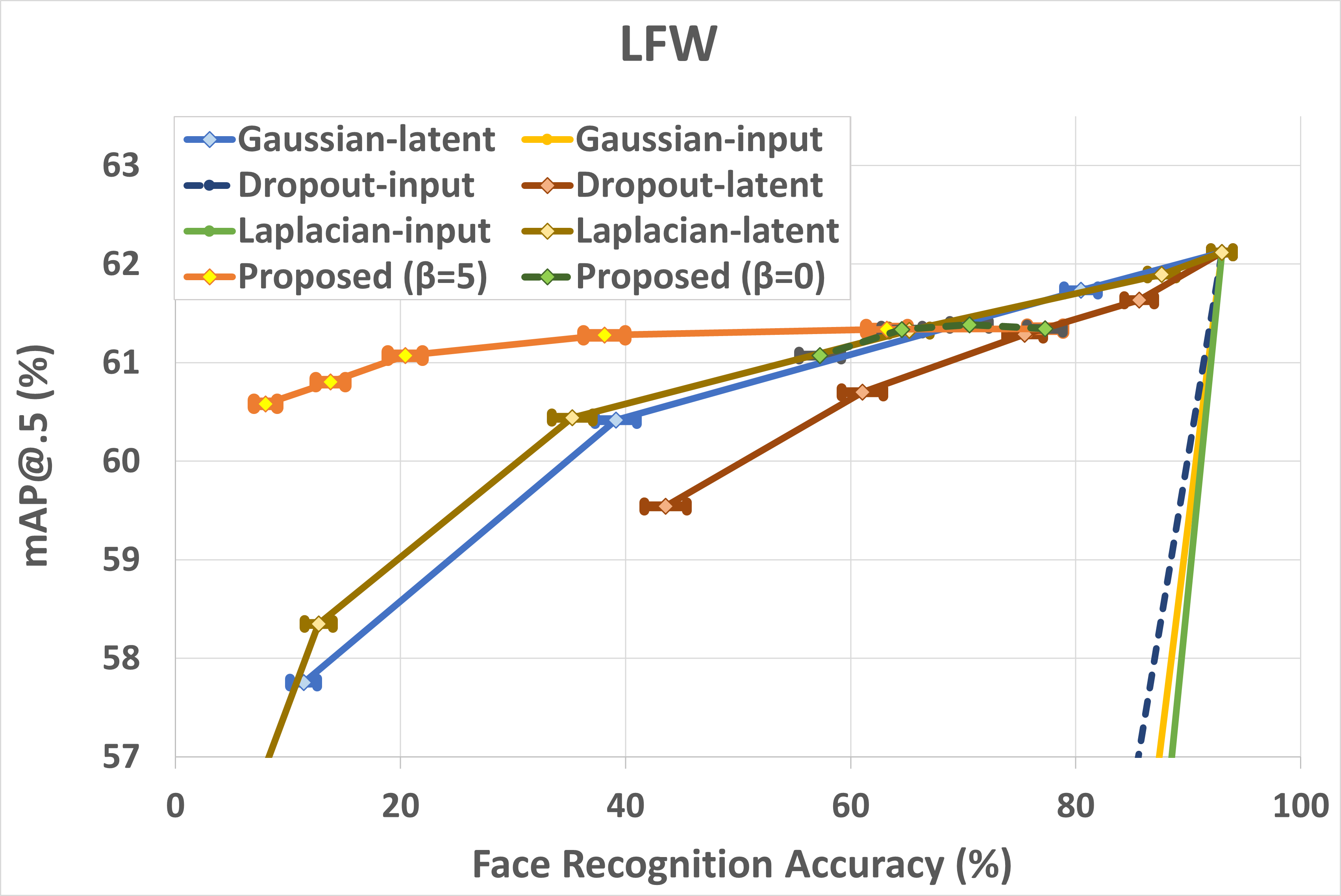}
        \label{subfig:wo_cmprs_face-acc}
    } 
    \hfill
    \subfloat[]{
        \centering
        \includegraphics[width=0.85\columnwidth]{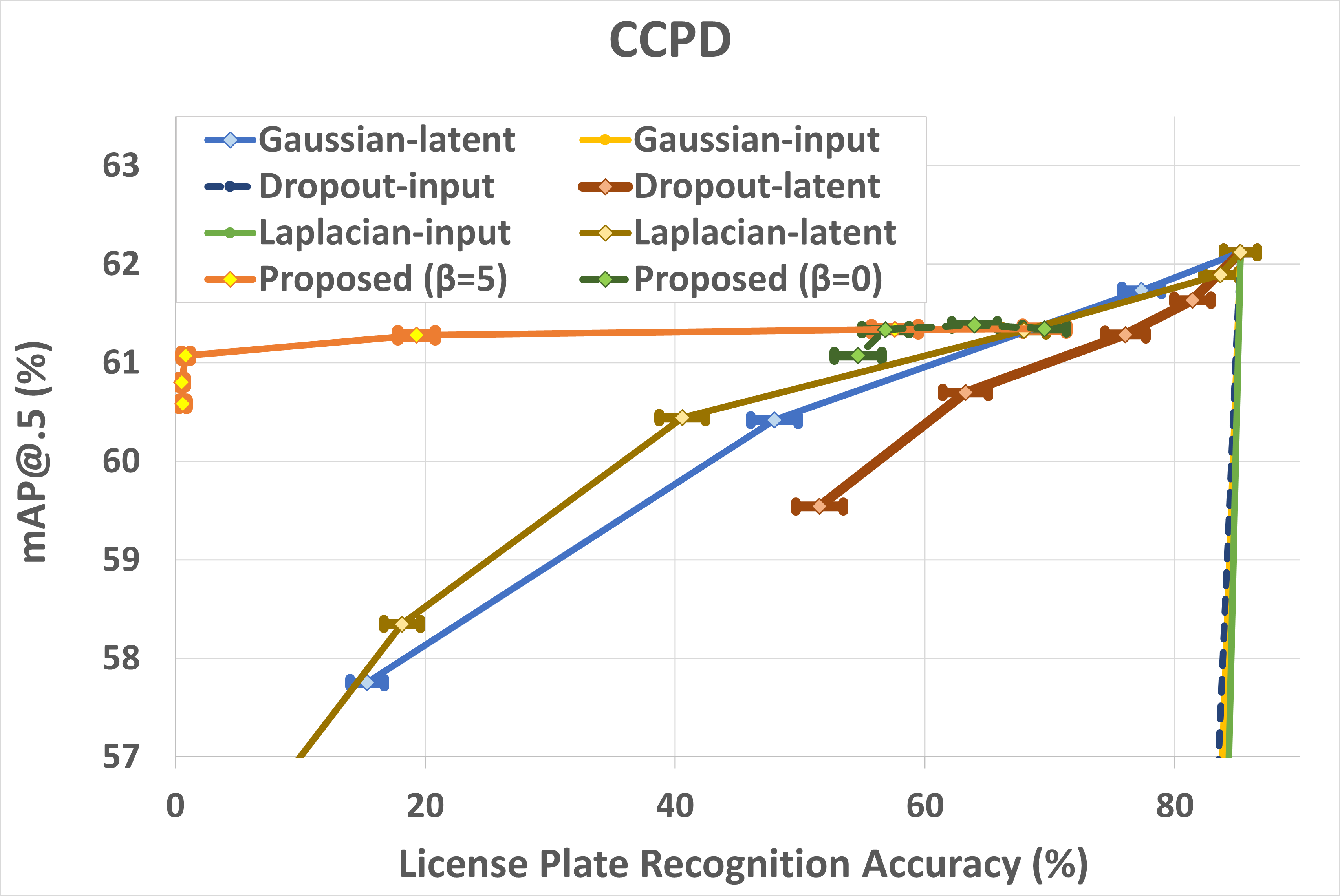}
        \label{subfig:wo_cmprs_license-acc}
    } 
    \caption{Privacy-Utility curves illustrating resistance to \ac{MIA}; bars around the curves' center points in (b) and (c) represent the $99.9\%$ confidence interval.}
    \label{fig:charts_wo_cmprs}
    \vspace{-0.3cm}
\end{figure}

Next, we present a comparison between the resistance of our proposed approach to \ac{MIA} and well-known defense strategies, such as adding Gaussian or Laplacian noise~\cite{diff_privacy_iot, not_just_privacy} and nullification (Dropout)~\cite{not_just_privacy, mia_2021}. These manipulations are applied with different intensities to either the input or latent features of YOLOv5m at layer 5 to obtain the corresponding Privacy-Utility curves. In those curves, the YOLO model is not fine-tuned to adapt to the manipulations, which ensures a fair comparison with our proposed method, wherein the YOLO model also remains fixed. Here, utility is measured by the accuracy of object detection on the COCO-2017 validation set (mean Average Precision, mAP, at IoU threshold of 0.5, shown on the vertical axis), and privacy is quantified in three different ways along the horizontal axis: PSNR of the recovered images on the COCO-2017 validation set in Fig.~\ref{subfig:wo_cmprs_psnr-acc}, face recognition accuracy on the LFW dataset in Fig.~\ref{subfig:wo_cmprs_face-acc}, and license plate recognition accuracy on the CCPD dataset in Fig.~\ref{subfig:wo_cmprs_license-acc}. Our autoencoder was trained with $w_{cmprs}=0$ and a set of values for $w_{rec}\in\{0.0, 0.1, 0.5, 0.8, 1.2, 2.0\}$ to plot the orange Privacy-Utility curves, where $\beta=5$, and $w_{rec}\in\{0.0, 0.1, 0.5, 2.0\}$ to plot the dashed dark green curves, where $\beta=0$, shown in Fig.~\ref{fig:charts_wo_cmprs}.

The bars around the center points (means) of the curves represent the $99.9\%$ confidence interval obtained from the normal distribution \cite{Altman_Machins_2013}. These intervals are calculated based on the sample mean and sample standard deviation on the LFW and CCPD datasets with $7,484$ and $5,000$ data points, respectively. According to Fig.~\ref{fig:charts_wo_cmprs}, the three leftmost points on our Privacy-Utility curves, corresponding to $w_{rec}\in\{0.8, 1.2, 2.0\}$, provide excellent privacy, with face and license plate recognition accuracies below $21\%$ and almost $0\%$, respectively. This level of privacy is achieved while sacrificing less than $1\%$ precision in the object detection task. The large gap between the confidence intervals of our method and those of the benchmark methods indicates that the difference in privacy between our approach and others at the same object detection accuracy is correspondingly large and statistically significant.

The comparison of the curves corresponding to $\beta=5$ and $\beta=0$ in Fig.~\ref{fig:charts_wo_cmprs} illustrates the efficacy of the edge-centric loss term proposed in~(\ref{eq:L_rec}). Although the reconstruction loss is not designed to directly target face and license plate recognition performance, our proposed method with $\beta=5$ is able to reduce the face and license plate recognition accuracy by approximately $37\%$ and $53\%$, respectively, at the same object detection accuracy of mAP@.5=$61\%$, compared to the case without the edge-centric loss ($\beta=0$). Therefore, the edge-centric loss term plays a critical role in eliminating task-irrelevant information and improving privacy.

\subsection{Compression Efficiency}
\label{subsec:compression}
\begin{figure}[!t]
    \centering
    \subfloat[]{
        \centering
        \includegraphics[width=0.85\columnwidth]{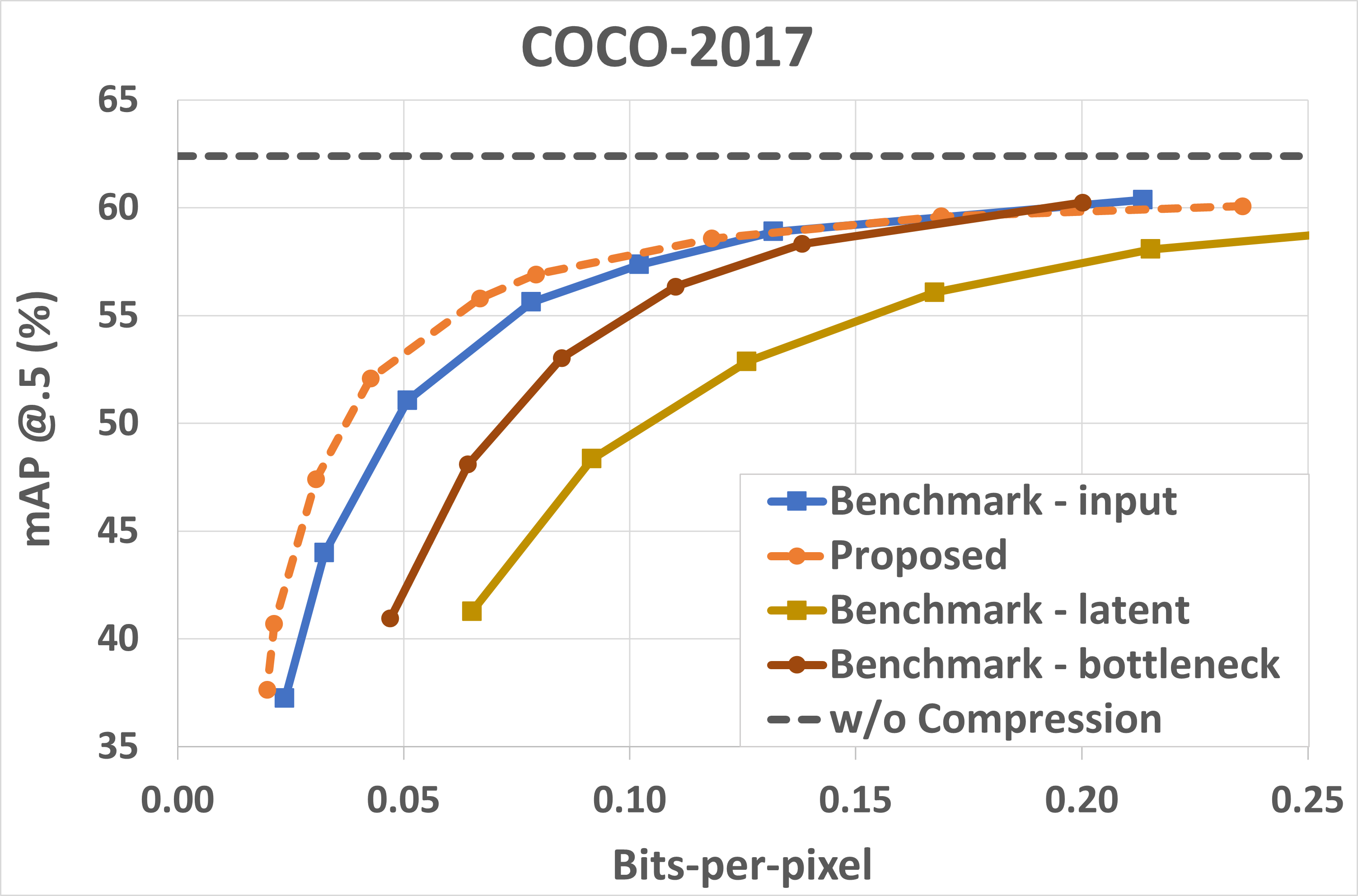}
        \label{subfig:cmprs_rate-acc}
    } 
    \hfill
    \subfloat[]{
        \centering
        \includegraphics[width=0.85\columnwidth]{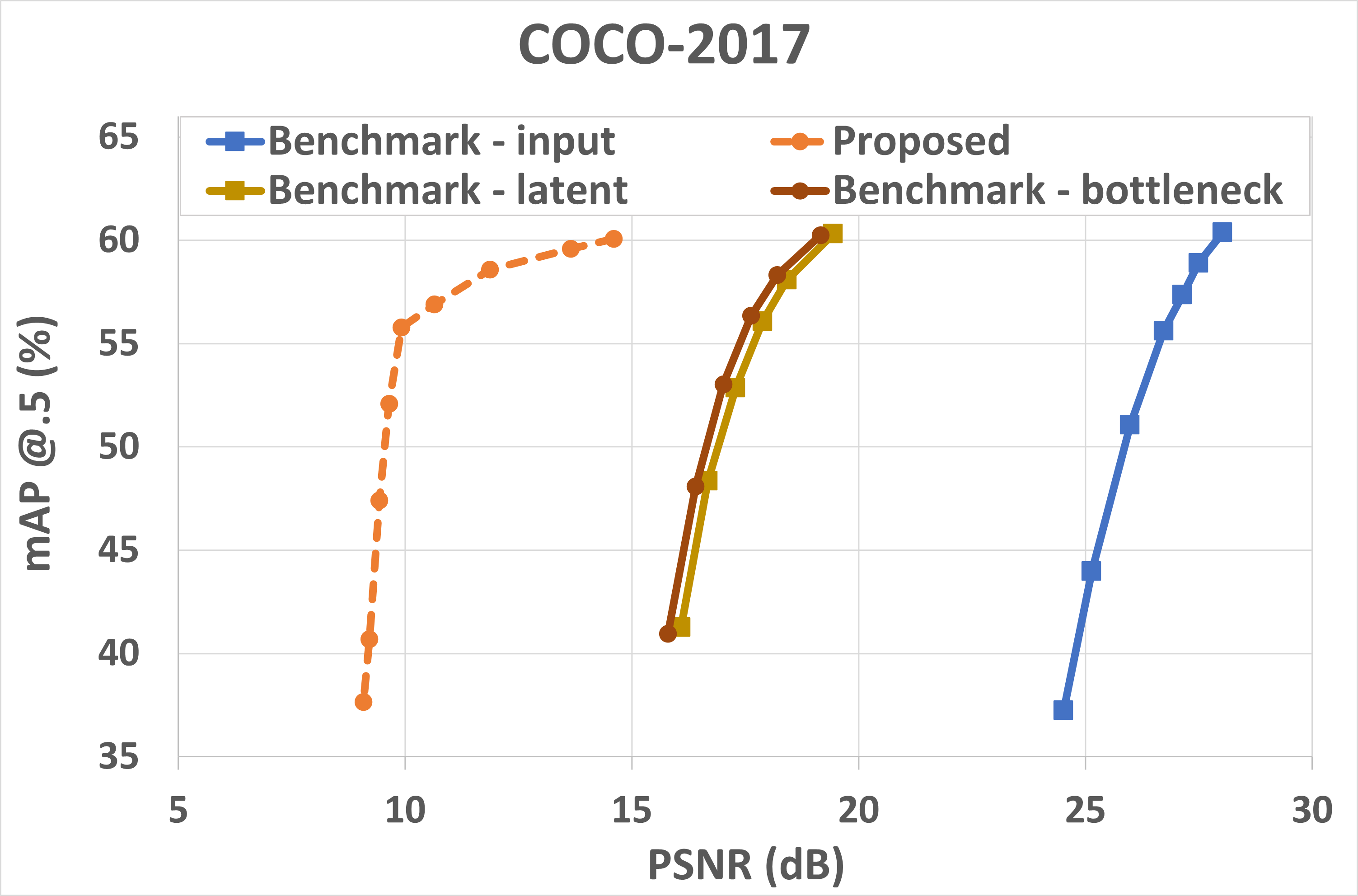}
        \label{subfig:cmprs_psnr-acc}
    } 
    \hfill
    \subfloat[]{
        \centering
        \includegraphics[width=0.85\columnwidth]{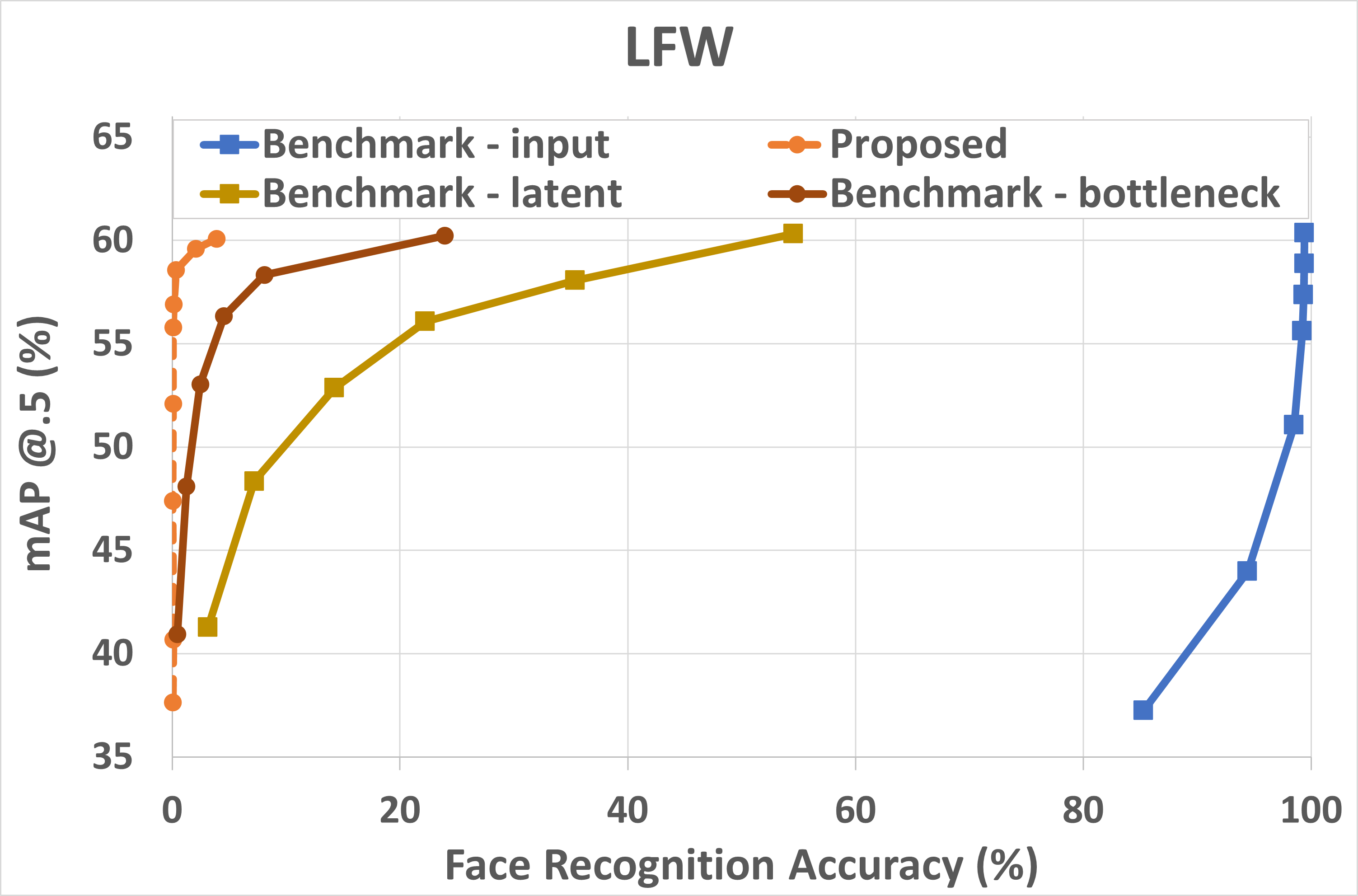}
        \label{subfig:cmprs_face-acc}
    } 
    \hfill
    \subfloat[]{
        \centering
        \includegraphics[width=0.85\columnwidth]{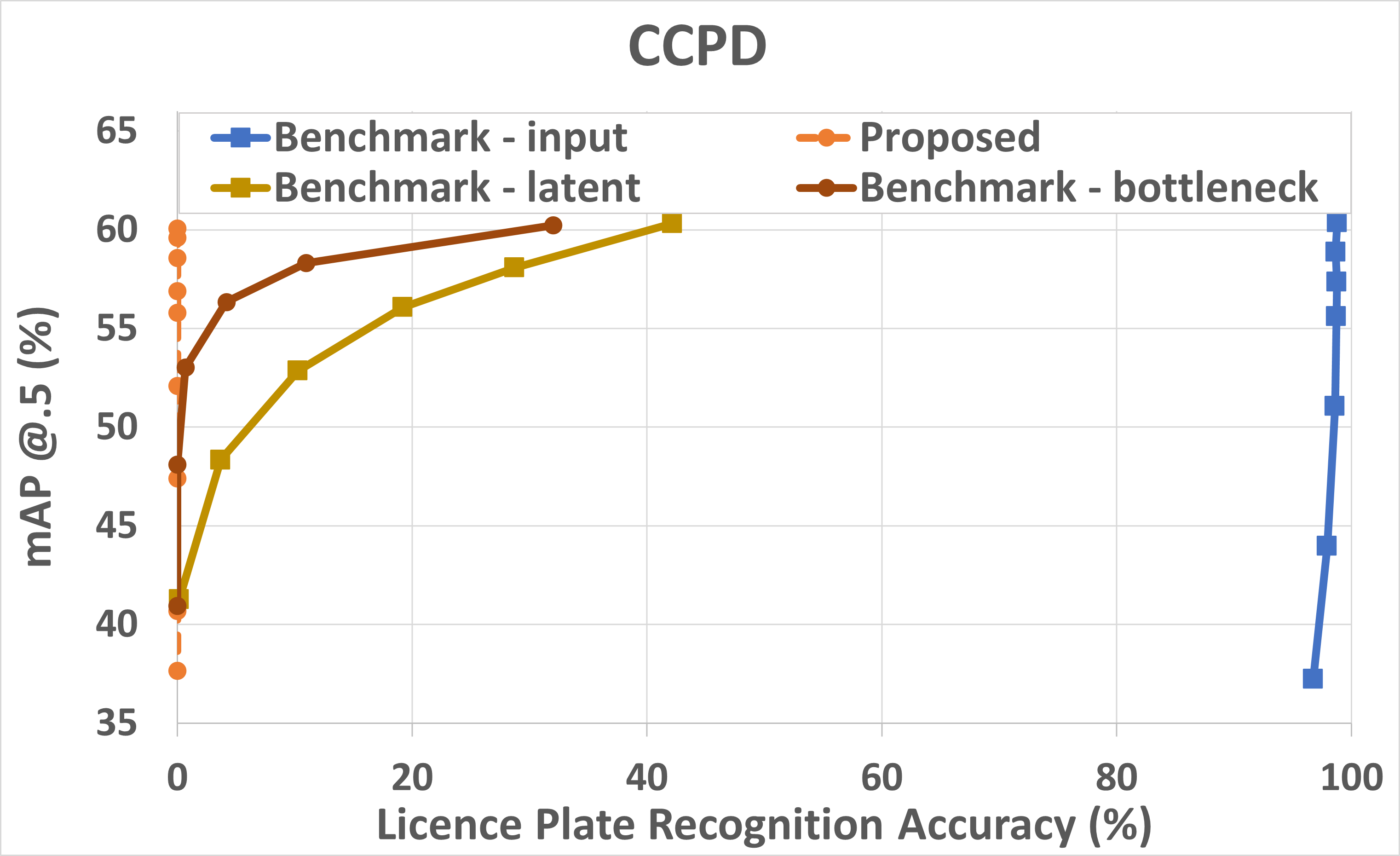}
        \label{subfig:cmprs_license-acc}
    } 
    \caption{(a) Rate-Utility curves;  (b), (c), (d) Privacy-Utility curves}
    \label{fig:charts_cmprs}
\end{figure}

In this section, we examine compression efficiency in addition to object detection and privacy. For this purpose, the VVC codec is included in all evaluation pipelines in Fig.~\ref{fig:benchmarks}.
For our approach, the key hyperparameters for training the autoencoder are $w_{rec}$ and $w_{cmprs}$ through which the amount of existing private information in the bottleneck and the bitrate of the coded bitstream can be controlled. We trained autoencoders with different values of $w_{rec} \in \{1.0, 1.5, 2.0\}$ and $w_{cmprs} \in \{1,2,3,4\}$. Note that, according to Fig.~\ref{fig:charts_wo_cmprs}, $1 \le w_{rec} \le 2 $ can offer suitable privacy protection. The bottleneck feature channels of each obtained autoencoder were compressed using the VVC codec with various \ac{QP} values. Therefore, each triplet $(w_{rec}, w_{cmprs}, QP)$ corresponds to a specific point in the Rate-Utility plane. The Pareto front of these points represents the performance of our proposed method alongside the benchmarks in Fig.~\ref{fig:charts_cmprs}.

For the benchmark methods, the curves are obtained by varying the \ac{QP} of the VVC codec. Fig.~\ref{subfig:cmprs_rate-acc} shows the Rate-Utility curves on the COCO-2017 validation set, where the rate is shown as the average number of bits per pixel (bpp), and the utility 
is mAP at the IoU threshold of 0.5, as before. Compressing YOLOv5m features directly (Benchmark-latent) shows the worst performance, because there are many channels (192) and the VVC codec is not tailored to coding features directly. Reducing the number of channels via autoencoder (Benchmark-bottleneck) improves the performance, but the features are still not VVC-friendly. Therefore, coding the input directly (Benchmark-input) shows better performance. However, the proposed approach creates VVC-friendly features at the autoencoder's bottleneck via the compressibility loss $L_{cmprs}$ in~(\ref{eq:compressibility_loss}), thereby surpassing the performance of Benchmark-input.

\begin{table}
    \begin{center}
    \caption{Bj{\o}ntegaard-Delta (BD) values with respect to ``Benchmark-input''}
    \begin{tabular}{|c|c|c|c|}
    \hline
         &  \textbf{BD-Rate} $\downarrow$  &  \textbf{BD-mAP} $\uparrow$ &  \textbf{BD-PSNR} $\downarrow$\\
    \hline\hline
    \textbf{Benchmark-input}  &  $0.0\%$  &  $0.0$  &  $0.00$~dB \\
    \hline
    \textbf{Benchmark-latent}  &  $110.5\%$  &  $-6.6$  &  $-8.90$~dB \\
    \hline
    \textbf{Benchmark-bottleneck}  &  $43.7\%$  &  $-3.1$  &  $-9.17$~dB \\
    \hline
    \textbf{Proposed}  &  $-19.4\%$  &  $1.9$  &  $-15.97$~dB \\
    \hline
    \end{tabular}
    \label{tab:BD}
    \end{center}
    \vspace{-0.7cm}
\end{table}

Figs.~\ref{subfig:cmprs_psnr-acc}-\ref{subfig:cmprs_license-acc} show the Privacy-Utility curves like those shown in Fig.~\ref{fig:charts_wo_cmprs}, but this time different points on the curves are obtained by varying the compression rate (via \ac{QP}) rather than changing the parameters of the defense mechanism. The results collectively show that our proposed method outperforms the benchmarks in various Privacy-Utility scenarios. One interesting point to note here is the performance of Benchmark-input, which is based on VVC coding of the input image directly. As shown in Fig.~\ref{subfig:cmprs_rate-acc}, this approach was the second best in terms of Rate-Utility, behind our method, which was not surprising because VVC-Intra is optimized for coding images. However, in terms of Privacy-Utility in Figs.~\ref{subfig:cmprs_psnr-acc}-\ref{subfig:cmprs_license-acc}, it is now significantly worse than other benchmarks. The likely reason is that compression artifacts introduced by VVC hurt object detection more than they affect face or license plate recognition. By operating on YOLOv5m features rather than input itself, the other two benchmarks and the proposed method manage to reduce the impact of compression on object detection performance.

We use Bj{\o}ntegaard-Delta (BD) values~\cite{bjontegaard2001calculation} to summarize the difference between various curves in Fig.~\ref{fig:charts_cmprs}. Table~\ref{tab:BD} shows BD-Rate and BD-mAP\footnote{Calculated based on the MPEG-VCM reporting template~\cite{mpeg_vcm_report}\label{foot:mpeg}} for the curves in Fig.~\ref{subfig:cmprs_rate-acc}, as well as the BD-PSNR for the curves shown in Fig.~\ref{subfig:cmprs_psnr-acc}. In the table, the symbol $\uparrow$ means the higher the better, while $\downarrow$ means the lower the better.
The BD-PSNR values indicate that the adversarially trained autoencoder can effectively reduce the PSNR of recovered images by $15.97$~dB and $7.07$~dB on average compared to Benchmark-input and Benchmark-latent, respectively. Moreover, a $19.4\%$ bit saving is achieved relative to Benchmark-input.

\subsection{Complexity}
As demonstrated in~\cite{kang2017neurosurgeon} and mentioned earlier in the text, collaborative inference has the potential to reduce overall latency compared to ``cloud-only'' solutions where input is directly compressed and transmitted to the cloud. We find that this is also true for the proposed system. Table~\ref{tab:time} shows the average execution time and the number of trainable parameters of different components of our proposed system, in the rows labeled ``Proposed.'' The bottom of the table, the rows labeled ``Input,'' is associated with the ``Benchmark-input'' method where input is compressed directly. For all components except the VVC codec,\footnote{VVenC~\cite{VVenC}, which was only able to run on CPU.} we show execution times on both the CPU and GPU. The CPU used for this experiment was AMD Ryzen 9 5900HS (3.30 GHz) and the GPU was NVIDIA GP102 TITAN X.

\begin{table}
    \begin{center}
    \caption{Average execution time (in milliseconds per input image) on the COCO-2017 validation set and the number of parameters for different components of the system}
    \vspace{-0.2cm}
    \begin{tabular}{|c|c|c|c|c|}
    \hline
    \multicolumn{2}{|c|}{\multirow{2}{*}{System component}} &
    \multicolumn{2}{c|}{Execution time (ms)} & \multirow{2}{*}{\textbf{\# Parameters}} \\
    \cline{3-4}
    \multicolumn{2}{|c|}{}  &  \textbf{CPU} &  \textbf{GPU} & {} \\
    \Xhline{3\arrayrulewidth}
    \multirow{7}{*}{\rotatebox[origin=c]{90}{\textbf{Proposed}}}
    & {YOLOv5m front-end}          &  $120.1$  &  $10.5$ & $723,168$ \\
    \cline{2-5}
    & {Autoencoder-AE}             &  $29.3$  &  $2.6$ & $590,080$ \\
    \cline{2-5}
    & {VVC bottleneck encoding}    &  \multicolumn{2}{c|}{$1481.1$} & - \\
    \cline{2-5}
    & {VVC bottleneck decoding}    &  \multicolumn{2}{c|}{$797.7$} & - \\
    \cline{2-5}
    & {Autoencoder-AD}             &  $27.7$  &  $1.6$ & $590,080$ \\
    \cline{2-5}
    & {YOLOv5m back-end}           &  $184.8$  &  $26.7$ & $20,467,389$ \\
    \cline{2-5}
    & \textbf{Total}   &  $\mathbf{2640.7}$  &  $\mathbf{2320.2}$ & $\mathbf{22,370,717}$ \\
    \Xhline{3\arrayrulewidth}

    \multirow{4}{*}{\rotatebox[origin=c]{90}{\textbf{Input}}}
    & {VVC input encoding}      &  \multicolumn{2}{c|}{$1833.4$} & - \\
    \cline{2-5}
    & {VVC input decoding}      &  \multicolumn{2}{c|}{$843.3$} & - \\
    \cline{2-5}
    & {YOLOv5m}                 &  $293.8$  &  $48.7$ & $21,190,557$ \\
    \cline{2-5}
    & \textbf{Total}  &  $\mathbf{2970.5}$  &  $\mathbf{2725.4}$ & $\mathbf{21,190,557}$ \\
    \hline
    
    \end{tabular}
    \label{tab:time}
    \end{center}
    \vspace{-0.3cm}
\end{table}

As seen in the table, most of the processing time is spent on VVC encoding, whereas the proposed autoencoder consumes the smallest portion, demonstrating its low complexity. The autoencoder's number of parameters is also relatively small, accounting for less than $6\%$ of the YOLO number of parameters. The proposed autoencoder generates a feature tensor that is tiled into a grayscale-like image of the same resolution as the input image. However, because the input image has 3 color channels, its encoding using VVC takes longer than the encoding of the tiled feature tensor. And the time difference ($1833.4-1481.1=352.3$ ms) is enough to compensate for running the YOLOv5m front-end and the autoencoder, even on a CPU ($120.1+29.3=149.4$ ms).

Simpler edge devices usually lack a GPU and therefore have to perform computations on a CPU. Hence, one practical realization of the system would be to run edge-side computations on a CPU and cloud-side computations on a GPU. In this case, the overall latency of our proposed method becomes $2456.5$~ms, which is around $10\%$ less than the time required when the input is compressed directly, which is $2725.4$~ms.

The above analysis was focused on the execution time of various components. But, the actual end-to-end inference latency also depends on the communication time, which is the time it takes for the encoded bitstream to be transmitted from the edge device to the cloud. As shown in Table~\ref{tab:BD}, at the same object detection accuracy, the proposed method provides BD-Rate savings of $19.4\%$, meaning that it produces, on average, a $19.4\%$ smaller bitstream. For a given transmission rate (in bits/second), such a bitstream would take $19\%$ less time to be transmitted compared to the Benchmark-input bitstream. Hence, the actual time savings in practice could be even higher than the $10\%$ achieved on execution time only.

\subsection{Ablation Study}
\label{subsec:ablation}

\begin{figure}[!t]
    \centering
    \subfloat[]{
        \centering
        \includegraphics[width=0.8\columnwidth]{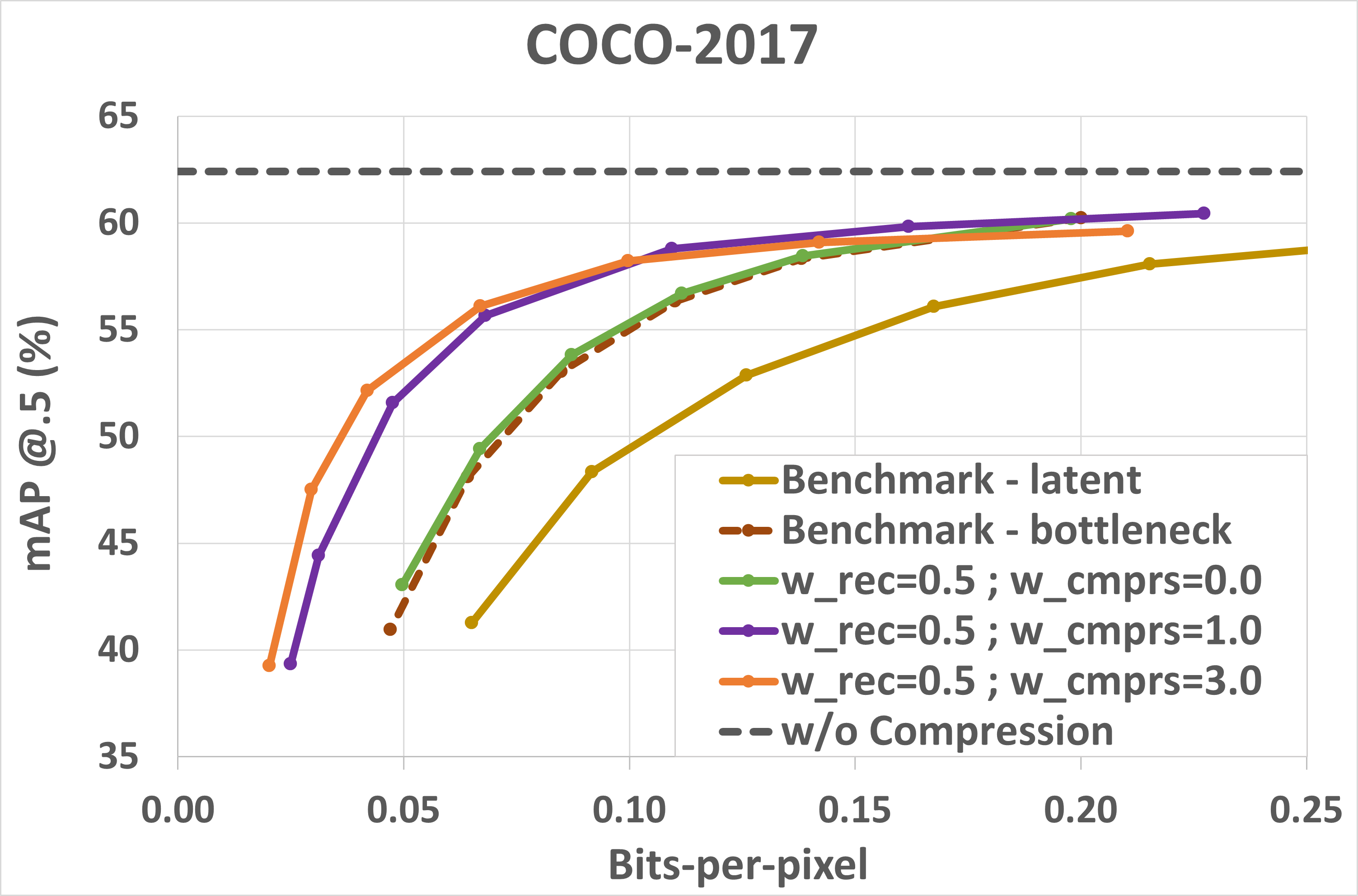}
        \label{subfig:ablate_rate-acc}
    } 
    \hfill
    \subfloat[]{
        \centering
        \includegraphics[width=0.8\columnwidth]{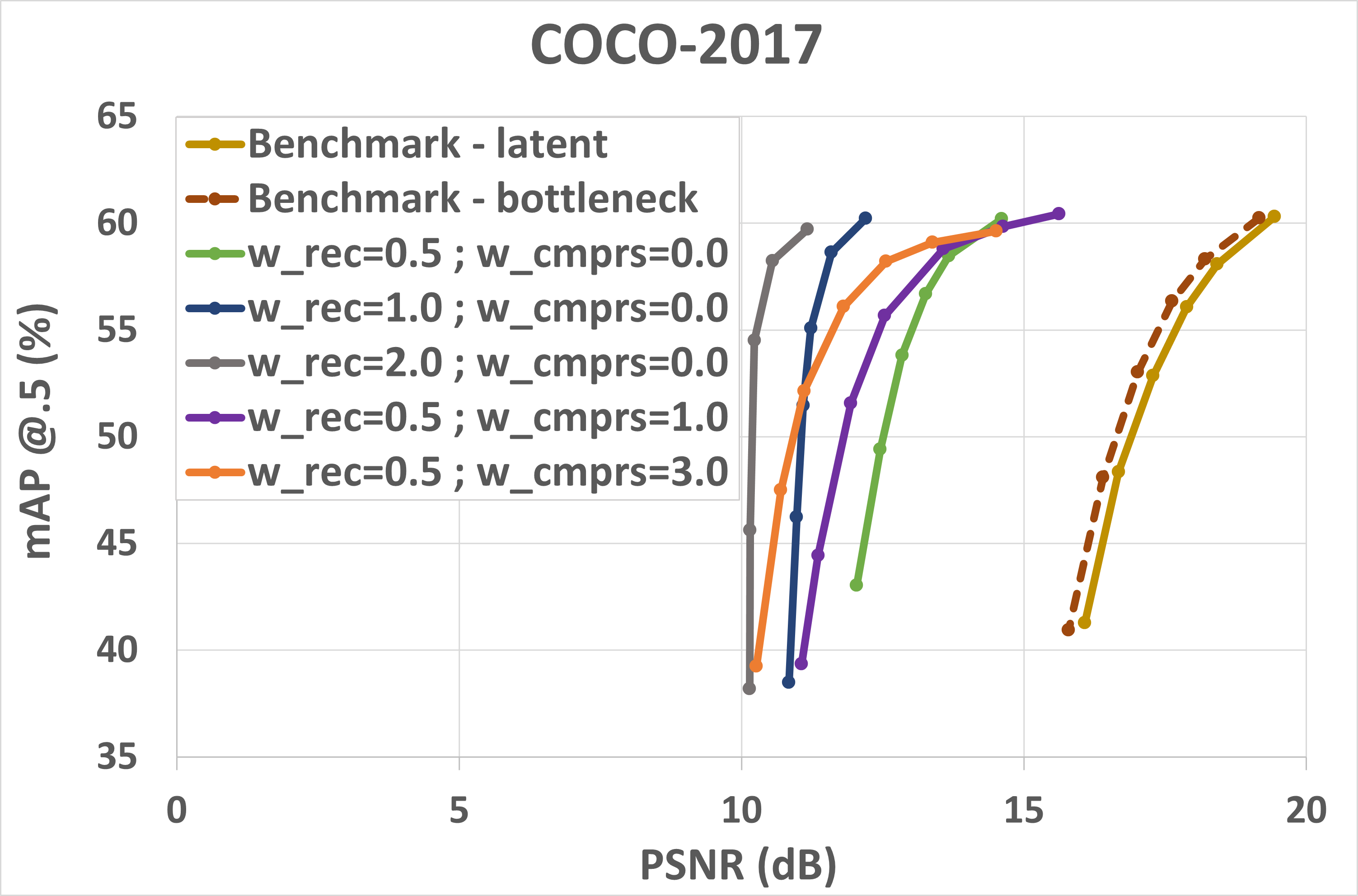}
        \label{subfig:ablate_psnr-acc}
    } 
    \caption{(a) Rate-Utility curves  (b) Privacy-Utility curves}
    \label{fig:ablate_charts}
    \vspace{-0.3cm}
\end{figure}

In this section, we analyze the impact of $w_{rec}$ and $w_{cmprs}$ on our system's performance.
Fig.~\ref{fig:ablate_charts} shows the Rate-Utility and Privacy-Utility curves for some of these values, with their corresponding BD values with respect to Benchmark-latent given in Table~\ref{tab:BD_ablate}. According to these results, Benchmark-bottleneck reduces the bitrate by $32\%$ compared to coding the original latent space of the YOLOv5m model. Thus, part of the bitrate reduction in the proposed system is attributed to the dimensionality reduction of the latent space. However, compressibility loss also plays a crucial role in minimizing the bitrate, which is seen when changing $w_{cmprs}$.

Another interesting point is that the average PSNR difference of the images recovered by \ac{MIA} from Benchmark-bottleneck and Benchmark-latent is negligible ($0.27$~dB). Hence, feature dimensionality alone is not sufficient to disrupt \ac{MIA}. On the other hand, adversarially trained autoencoders with $w_{rec}>0$ provide a more significant reduction in quality of \ac{MIA}-generated images, especially when $w_{rec}$ increases. 

It is also interesting to note that $w_{rec}$ and $w_{cmprs}$ provide somewhat independent ``knobs'' to control the quality of \ac{MIA}-generated images and bitrate, respectively. For instance, increasing $w_{rec}$ from $0.5$ to $1.0$ or $2.0$ (with $w_{cmprs}=0$) minimally alters the average bitrate, yet leads to an average PSNR drop of about $2.3$~dB in the quality of \ac{MIA}-generated images. Similarly, increasing $w_{cmprs}$ from $0.0$ to $3.0$ (with $w_{rec}=0.5$) has a relatively minor impact on the quality of \ac{MIA}-generated images (especially at high bitrates) compared to the effect of $w_{rec}$, yet it leads to an additional $30\%$ bitrate reduction relative to Benchmark-latent.

\begin{table}
    \begin{center}
    \caption{Bj{\o}ntegaard-Delta (BD) values with respect to ``Benchmark-latent''}
    \vspace{-0.2cm}
    \begin{tabular}{|c|c|c|c|}
    \hline
         &  \textbf{BD-Rate} $\downarrow$  &  \textbf{BD-mAP} $\uparrow$ &  \textbf{BD-PSNR} $\downarrow$\\
    \hline\hline
    Benchmark-latent  &  $0.0\%$  &  $0.0$  &  $0.00$~dB \\
    \hline
    Benchmark-bottleneck  &  $-32.0\%$  &  $5.0$  &  $-0.27$~dB \\
    \hline
    $w_{rec}=0.5, w_{cmprs}=0.0$   &  $-34.1\%$  &  $5.2$  &  $-4.47$~dB \\
    \hline
    ${w_{rec}=1.0, w_{cmprs}=0.0}$  &  $-36.7\%$  &  $5.5$  &  $-5.95$~dB \\
    \hline
    ${w_{rec}=2.0, w_{cmprs}=0.0}$  &  $-32.7\%$  &  $5.0$  &  $-6.79$~dB \\
    \hline
    ${w_{rec}=0.5, w_{cmprs}=1.0}$  &  $-57.2\%$  &  $7.1$  &  $-5.05$~dB \\
    \hline
    ${w_{rec}=0.5, w_{cmprs}=3.0}$  &  $-63.3\%$  &  $7.2$  &  $-5.91$~dB \\
    \hline
    \end{tabular}
    \label{tab:BD_ablate}
    \end{center}
    \vspace{-0.4cm}
\end{table}

\section{Conclusion}
\label{sec:conclusion}
In this paper, we presented a novel feature coding scheme for collaborative object detection with strong resistance to model inversion attack (MIA). This was achieved via an autoencoder that transforms the features of the YOLOv5m object detection model into a lower-dimensional space while maintaining object detection accuracy, and improving compressibility as well as resistance to \ac{MIA}. 
Moreover, the proposed approach enables lower end-to-end inference latency compared to encoding the input image directly.
The proposed approach is plug-and-play in the sense that it utilizes a standard widely available codec (VVC) and does not require retraining the object detector. It is also lightweight and suitable for edge-device deployment. Nonetheless, using a more complex autoencoder and training it jointly with the object detection front-end and/or back-end could enhance the model's flexibility in eliminating private and retaining task-relevant information, while improving feature compressibility. This approach may be a promising direction for future research in this area.

In addition to the conventional evaluation of \ac{MIA} resistance in terms of the quality of \ac{MIA}-generated images, the proposed system was also evaluated on measures that are more directly related to privacy, such as face and license plate recognition accuracy, showing superior performance compared to benchmarks in each case.


\bibliographystyle{ieeetr}
\bibliography{reference}

\begin{IEEEbiography}
[{\includegraphics[width=1in,height=1.25in,clip,keepaspectratio]{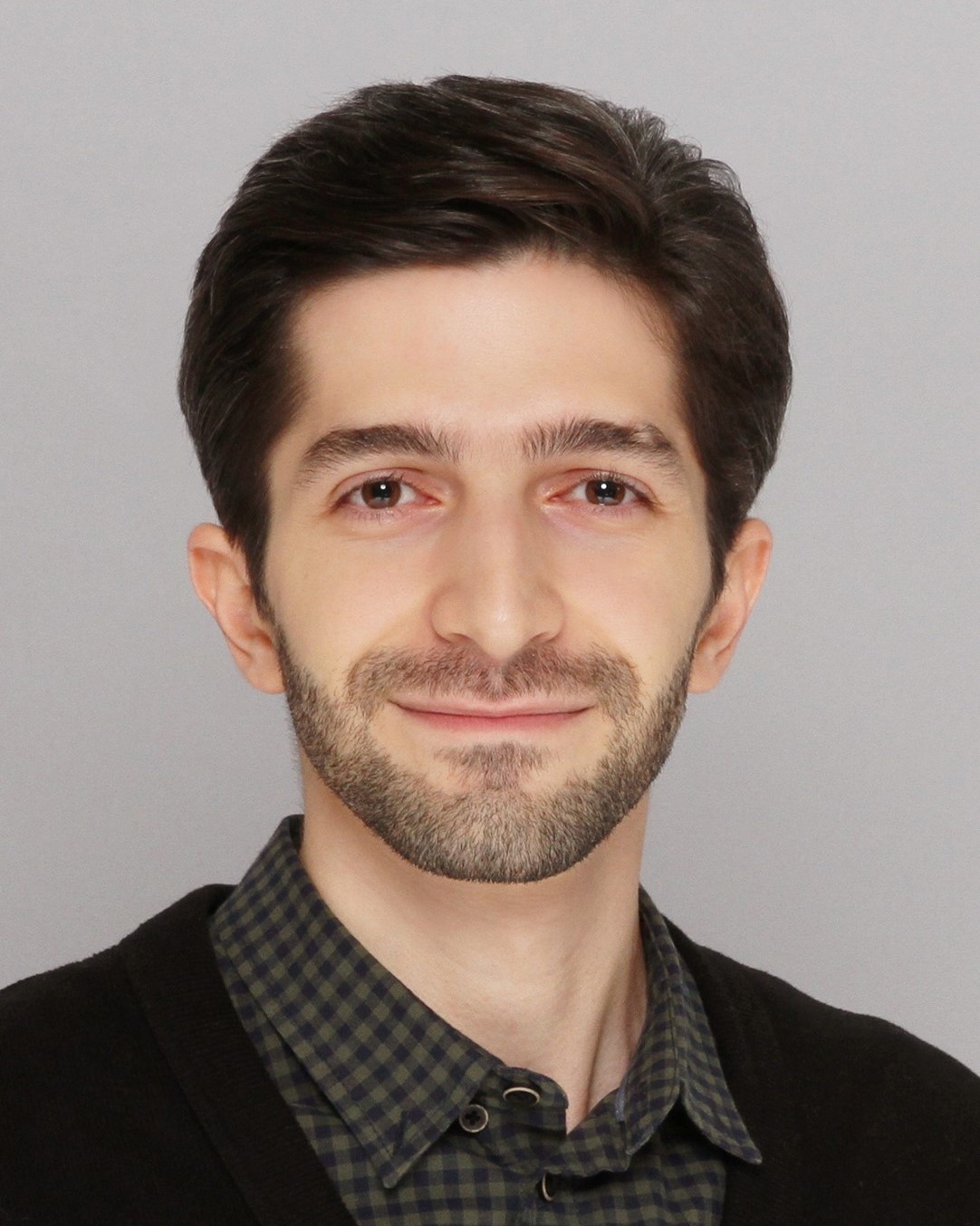}}]{Bardia Azizian}
received the B.Sc. degree in electrical engineering from the University of Tehran, Iran, in 2015, the M.Sc. degree in secure communication and cryptography from the Sharif University of Technology, Iran, in 2018, and the M.A.Sc. degree from Simon Fraser University, Canada, in 2023, where he is currently pursuing a Ph.D. degree. His research interests include deep learning and signal processing, with a focus on data compression and coding for multi-modal machines.
\end{IEEEbiography}

\begin{IEEEbiography}
[{\includegraphics[width=1in,height=1.25in,clip,keepaspectratio]{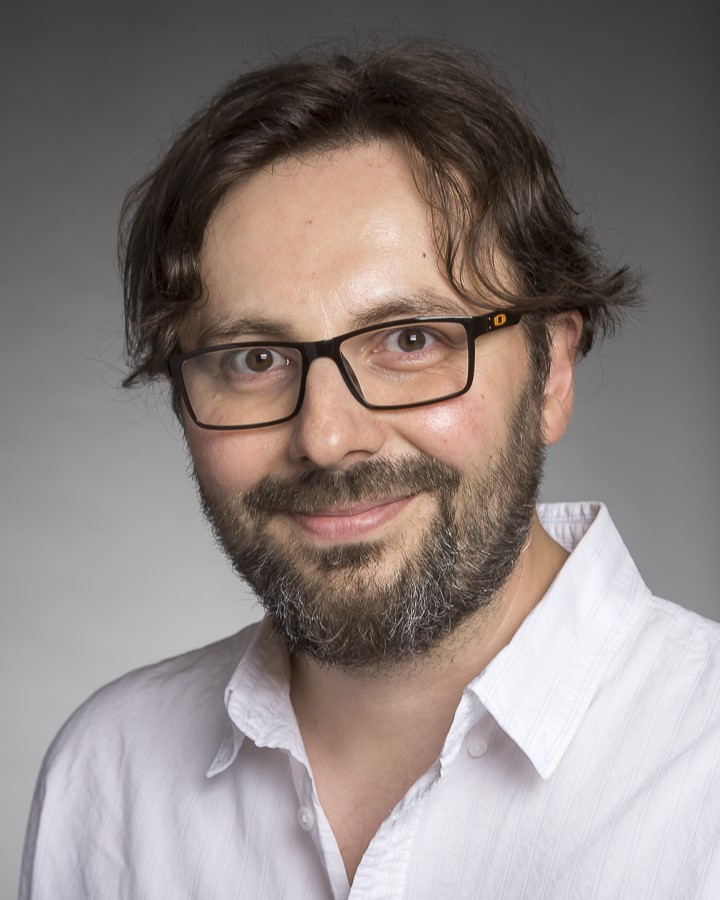}}]{Ivan V. Baji\'{c}}
(Senior Member, IEEE) is a Professor of engineering science and the Co-Director of the Multimedia Laboratory at Simon Fraser University, Burnaby, BC, Canada. His research interests include signal processing and machine learning, with applications in multimedia signal compression, processing, and collaborative intelligence. His group’s work has received several research awards including the 2023 IEEE TCSVT Best Paper Award, conference paper awards at ICME 2012, ICIP 2019, MMSP 2022, and ISCAS 2023; and other recognitions (e.g., paper award finalist, top n\%) at Asilomar, ICIP, ICME, ISBI, and CVPR. He is a Past Chair of the IEEE Multimedia Signal Processing Technical Committee. He was on the editorial boards of IEEE Transactions on Multimedia, IEEE Signal Processing Magazine, and Signal Processing: Image Communication, and is currently a Senior Area Editor of the IEEE Signal Processing Letters.
\end{IEEEbiography}

\end{document}